\documentclass[journal,twocolumn]{IEEEtran}
\usepackage{amsmath,amssymb,amsfonts}
\usepackage{algorithmic}
\usepackage{textcomp}
\usepackage{tabularx}

\usepackage[colorinlistoftodos]{todonotes}
\usepackage{enumitem}
\usepackage{multirow}
\usepackage{multicol}
\usepackage{graphicx}
\usepackage{bm}
\usepackage{colortbl}
\usepackage[english]{babel}
\usepackage{xspace}
\usepackage{multirow}
\usepackage{rotating}
\usepackage{fancyhdr}
\usepackage{booktabs}
\usepackage{underscore}
\usepackage{footmisc}
\usepackage{subfigure}
\usepackage{url}
\usepackage{algorithm}
\usepackage[noadjust]{cite}
\usepackage{makecell}
\usepackage{array}
\usepackage{comment}
\usepackage{lettrine}
\usepackage{color}
\usepackage{xspace}
\usepackage{rotating}
\definecolor{mycyan}{gray}{0.95}
\usepackage{threeparttable}
\newcommand{\mypara}[1]{{\vspace{0.05mm}\noindent\textbf{#1}}}

\newcommand{\projtitle}{\textsc{Hawk}\xspace}

\newcommand{\HIN}{\textsc{Hin}\xspace}

\newcommand{\modelbasic}{\textsc{MsGAT}\xspace}
\newcommand{\modelplus}{\textsc{MsGAT++}\xspace}

\newcommand\FIXME[1]{\textcolor{red}{FIX: }\textcolor{red}{#1}}

\def\BibTeX{{\rm B\kern-.05em{\sc i\kern-.025em b}\kern-.08em
    T\kern-.1667em\lower.7ex\hbox{E}\kern-.125emX}}
\begin{document}

\title{\projtitle: Rapid Android Malware Detection through Heterogeneous Graph Attention Networks}

\author{
Yiming~Hei, 
Renyu~Yang,~\IEEEmembership{Member,~IEEE},
Hao~Peng, 
Lihong~Wang,
Xiaolin~Xu,
Jianwei~Liu,
Hong~Liu,
Jie~Xu,~\IEEEmembership{Member,~IEEE},
Lichao~Sun
\IEEEcompsocitemizethanks{
\noindent \IEEEcompsocthanksitem Manuscript received Mar 2021, revised June 2021, accepted August 2021.
This work was supported by the NSFC Grants (62002007, U20B2053, 62073012 and 62072184), S\&T Program of Hebei Grant (20310101D), Fundamental Research Funds for the Central Universities, Project of Science and Technology Commitment of Shanghai Grant 20511106002, the UK EPSRC (EP/T01461X/1) and UK White Rose University Consortium, and Opening Project of Shanghai Trusted Industrial Control Platform. R.Yang would also appreciate the birth of Ruisi and numerous sleepless but encouraging nights with her when preparing this manuscript.
\textit{(Corresponding author: Hao Peng. \textit{R.Yang and Y.Hei are co-first authors with equal contribution.})}
\IEEEcompsocthanksitem Y.Hei, H.Peng and J.Liu are with the School of Cyber Science and Technology, Beihang University, Beijing 100083, China. Email:\{black, penghao, liujianwei\}@buaa.edu.cn.
\IEEEcompsocthanksitem R.Yang and J.Xu are with the School of Computing, University of Leeds, Leeds LS2 9JT, UK. Email: \{r.yang1, j.xu\}@leeds.ac.uk. 
\IEEEcompsocthanksitem L.Wang and X.Xu are with the National Computer Network Emergency Response Technical Team/Coordination Center of China, Beijing 100029, China. Email: \{wlh, xxl\}@isc.org.cn.
\IEEEcompsocthanksitem H.Liu is with the School of Computer Science and Software Engineering, East China Normal University, and with Shanghai Trusted Industrial Control Platform Co., Ltd., Shanghai 200062, China. Email: liuhong@ticpsh.com.
\IEEEcompsocthanksitem L.Sun is with the Department of Computer Science and Engineering, Lehigh University, Bethlehem, USA. Email: james.lichao.sun@gmail.com.
}
}

\markboth{IEEE TRANSACTIONS ON NEURAL NETWORKS AND LEARNING SYSTEMS,~Vol.~XXX, No.~XXX, XXX~2021}%
{Hei \MakeLowercase{\textit{et al.}}:\projtitle: HAWK: Rapid Android Malware Detection through Heterogeneous Graph Attention Networks}

\IEEEtitleabstractindextext{
\begin{abstract}
Android is undergoing unprecedented malicious threats daily, but the existing methods for malware detection often fail to cope with evolving camouflage in malware. To address this issue, we present \projtitle, a new malware detection framework for evolutionary Android applications. We model Android entities and behavioural relationships as a heterogeneous information network (\HIN), exploiting its rich semantic meta-structures for specifying implicit higher-order relationships. An incremental learning model is created to handle the applications that manifest dynamically, without the need for re-constructing the whole \HIN and the subsequent embedding model. The model can pinpoint rapidly the proximity between a new application and existing in-sample applications and aggregate their numerical embeddings under various semantics. Our experiments examine more than 80,860 malicious and 100,375 benign applications developed over a period of seven years, showing that \projtitle achieves the highest detection accuracy against baselines and takes only 3.5ms on average to detect an out-of-sample application, with the accelerated training time of 50x faster than the existing approach.
\end{abstract}

\begin{IEEEkeywords}
Android, malware detection, graph representation learning, HIN
\end{IEEEkeywords}
}

\maketitle
\IEEEdisplaynontitleabstractindextext

\section{Introduction}
\label{sec:intro}

\IEEEPARstart{W}{ITH} the highest market share worldwide on mobile devices, Android is experiencing unprecedented dependability issues. 
Due to Android's extensibility and openness of development, users are put at high risk of a variety of threats and illegal operations from malicious software, i.e., \textit{malware} including privacy violations, data leakage, advertisement spams, etc. 
Common Vulnerabilities and Exposures (CVE) reveals 414 Android vulnerabilities that can be easily attacked in realistic environments.
This phenomenon  calls for more reliable and accessible detection techniques.

Conventionally, Android Applications (Apps) are analyzed by either static 
analysis, through pre-determined signatures/semantic artifacts, or dynamic analysis through multi-level instrumentation~\cite{ye2017survey}. However, static analysis could become invalid by simple obfuscation, while dynamic analysis heavily depends on OS versions and the Android runtime, which is inherently cost-expensive and time-consuming. 
To tackle this, numerous machine-learning based detection techniques~\cite{2mclaughlin2017deep, li2018significant,hou2016droiddelver, dimjavsevic2015android,hou2016deep4maldroid, alzaylaee2020dl,wang2020deep} typically leverage feature engineering to extract key malware features and apply classification algorithms -- each app is represented as a vector -- to distinguish benign software from malicious software. Nevertheless, these approaches often fail to capture emerging malware that either conducts evolving camouflage and attack type or hides certain features deliberately\footnote{https://www.mcafee.com/blogs/other-blogs/mcafee-labs}. 
Hence, it is imperative to build an inductive and rapid mechanism for constantly capturing software evolution and detecting malware without heavily relying on domain-specific feature selection. 

Graph neural network (GNN), which is used to model the relationship between entities, is developing rapidly in both theoretical~\cite{kipf2016semi,velivckovic2017graph,PEGNN,Haar} and applied fields~\cite{peng2021streaming,peng2019hierarchical}. Heterogeneous information network (\HIN) ~\cite{sun2011pathsim,peng2021lime}, as a special case of graph neural network, has been widely adopted in many areas such as operating systems, Internet of Things and cyber-security by exploiting the abundant node and relational semantic information before embedding into representation vectors~\cite{fan2018gotcha,3hou2017hindroid,mgnet,wang2019attentional}. More specifically, in the context of malware detection, if $App_1$ and $App_2$ share permission \texttt{SEND\_SMS} while $App_2$ and $App_3$ share permission \texttt{READ\_SMS}, \HIN is able to capture the implicit semantic relationship between $App_1$ and $App_3$ that can be hardly achieved by feature engineering based approaches. \HIN-based modelling is even more meaningful because malware developers are extremely difficult to hide such implicit relationships~\cite{3hou2017hindroid}. While promising, \HIN is inherently concerned about static networks/graphs~\cite{ye2018aidroid}. The complication is, however, how to efficiently embed the \textit{out-of-sample} nodes (i.e., incoming nodes out of the established \HIN). Considering the continuous software updates and the huge volume of Apps, it is impossible to involve all Apps in the stage of \HIN construction and inefficient to re-construct the entire embedding model when new Apps are seen emerging. This drawback impedes the practicality and the scale this native technique can perform. Although AiDroid~\cite{ye2018aidroid} attempts to tackle this problem and represents each out-of-sample App with convolutional neural network (CNN)~\cite{simonyan2014very}, it requires heavily multiple convolution operations resulting in  non-negligible time inefficiency. 

In this paper, we present \projtitle, a novel Android malware detection framework with the aid of network representation learning model and \HIN to explore abundant but hidden semantic information among different Apps. In particular, we extract seven types of Android entities -- including App, permission, permission type, API, class, interface and \texttt{.so} file -- from the decompiled Android application package (APK) files and establish a \HIN mainly through transforming entities and their relationships into nodes and edges, respectively. We exploit rich semantic meta structures as the templates to define relation sequence between two entity types. 
This includes both meta path ~\cite{dong2017metapath2vec} and meta graph ~\cite{zhang2018metagraph2vec} that can specify the implicit relationships among heterogeneous entities. A certain meta structure corresponds to an adjacency matrix associated with a homogeneous graph. The graph only contains App nodes and is the target in the procedure of malware detection. 

At the core of \projtitle is the numerical embedding of all App entities that can be then fed into a binary classifier. In particular, \projtitle involves two distinct learning models for in-sample and out-of-sample nodes, respectively. To embed an in-sample  App, we propose \modelbasic, a meta structure guided graph attention network mechanism~\cite{vaswani2017attention} that incorporates its neighbors' embedding within any meta structure and integrates the embedding results of different meta structures into the final node embedding.   This design takes into account not only the informative connectivity of neighbor nodes but also the diverse semantic implications over different entity relationships. In addition, to efficiently embed an out-of-sample  App, we present \modelplus, a new incremental learning model upon \modelbasic to make good use of the embedding of certain existing nodes. Given a specific meta structure and its corresponding graph, our model firstly pinpoints a specific set of in-sample App nodes that are most similar to the target new node, before aggregating their embedding vectors to form the node embedding under this meta structure. 
Likewise, we entitle particular weights to individual embedding vector of each meta structure and aggregate them to obtain the final embedding. This incremental design can quickly calculate the embedding based on the established \HIN structures without re-learning the holistic embedding for all nodes, thereby significantly improving the training efficiency and model scalability. 

We demonstrate the effectiveness and efficiency of \projtitle based on 80,860 malicious and 100,375 benign Apps collected and decompiled across VirusShare, CICAndMal and Google AppStore. Experiments show that \projtitle outperforms all baselines in terms of accuracy and F1 score, indicating its effectiveness and suitability for malware detection at scale. It takes merely 3.5 milliseconds on average to detect
an out-of-sample App with accelerated training time of 50$\times$ against the native approach that rebuilds the \HIN and reruns the \modelbasic. To enable replication and foster research, we make \projtitle publicly available at: \texttt{\url{github.com/RingBDStack/HAWK}}. This paper makes the following contributions:

$\bullet$ It examines 200,000+ Android Apps and decompiled 180,000+ APKs, spanning over seven years across multiple open repositories. This discloses abundant data source to establish the \HIN and uncovers the hidden high-order semantic relationships among Apps (\S~\ref{sec:hinconstr}).
	
$\bullet$ It presents a meta-structure guided attention mechanism based on \HIN for node embedding, by fully exploiting neighbor nodes within and across meta structures (\S~\ref{sec:models:insample}). Experiments show the capture of semantics can support excellent forward and backward compatible detection capabilities.
    
$\bullet$ It proposes an incremental aggregation mechanism for rapidly learning the embedding of out-of-sample Apps, without compromising the quality of numerical embedding and detection effectiveness. (\S~\ref{sec:models:outofsample}).


\mypara{Organization.} \S~\ref{sec:background} depicts the motivation and outlines the system overview. \S~\ref{sec:hinconstr} discusses the procedure of feature engineering and data reshaping by leveraging \HIN while \S~\ref{sec:models} details the core techniques to tackle in-sample and out-of-sample malware detection. Experimental set-up and results are presented in \S~\ref{sec:exp_setup} and \S~\ref{sec:exp}. Related work is  discussed in \S~\ref{sec:secRework} before we conclude the paper and discuss the future work.

\vspace{-0.1mm}
\section{Background and Overview}
\label{sec:background}

\vspace{-0.1mm}
\subsection{Motivation and Problem Scope}
\label{sec:background:motivation} 
\vspace{-0.2mm}

The Android platform is increasingly exposed to various malicious threats and attacks. As malware detection for Android systems is a response-sensitive task, our work addresses two primary research challenges -- \textit{inductive capability} and \textit{detection rapidness}. Anomaly identification should allow for forecasting new applications that we have not seen (the so-called \textit{out-of-sample} Apps) and rapidly catch up the up-to-date malicious attacks and threats, particularly considering the vast diversity and rapid growth of emerging malicious software.

The detection procedure is typically regarded as a binary classification.  Formally, we aim to take as input features $\mathcal{X}$ of Android Apps and their previous labels (malicious/benign) $\mathcal{T}$ to predict the type $t$ of any target App either old or new. 
Unfortunately, the existing approaches for malware detection are inadequate in tackling inductive problems where new application is arbitrary and unseen beforehand. Most of prior work on network embedding~\cite{dong2017metapath2vec, wang2019heterogeneous,zhao2017meta,zhang2018metagraph2vec} are \textit{transductive}, i.e., if a new data point is added to the testing dataset, one has to thoroughly re-train the learning model. 
Hence, malware detection is in great need of a generic \textit{inductive} learning model where any new data would be predicted, based on an observed set of training set, without the need to re-run the whole learning algorithm from scratch. 

\begin{figure}[t]
\centerline{\includegraphics[width=0.345\textwidth]{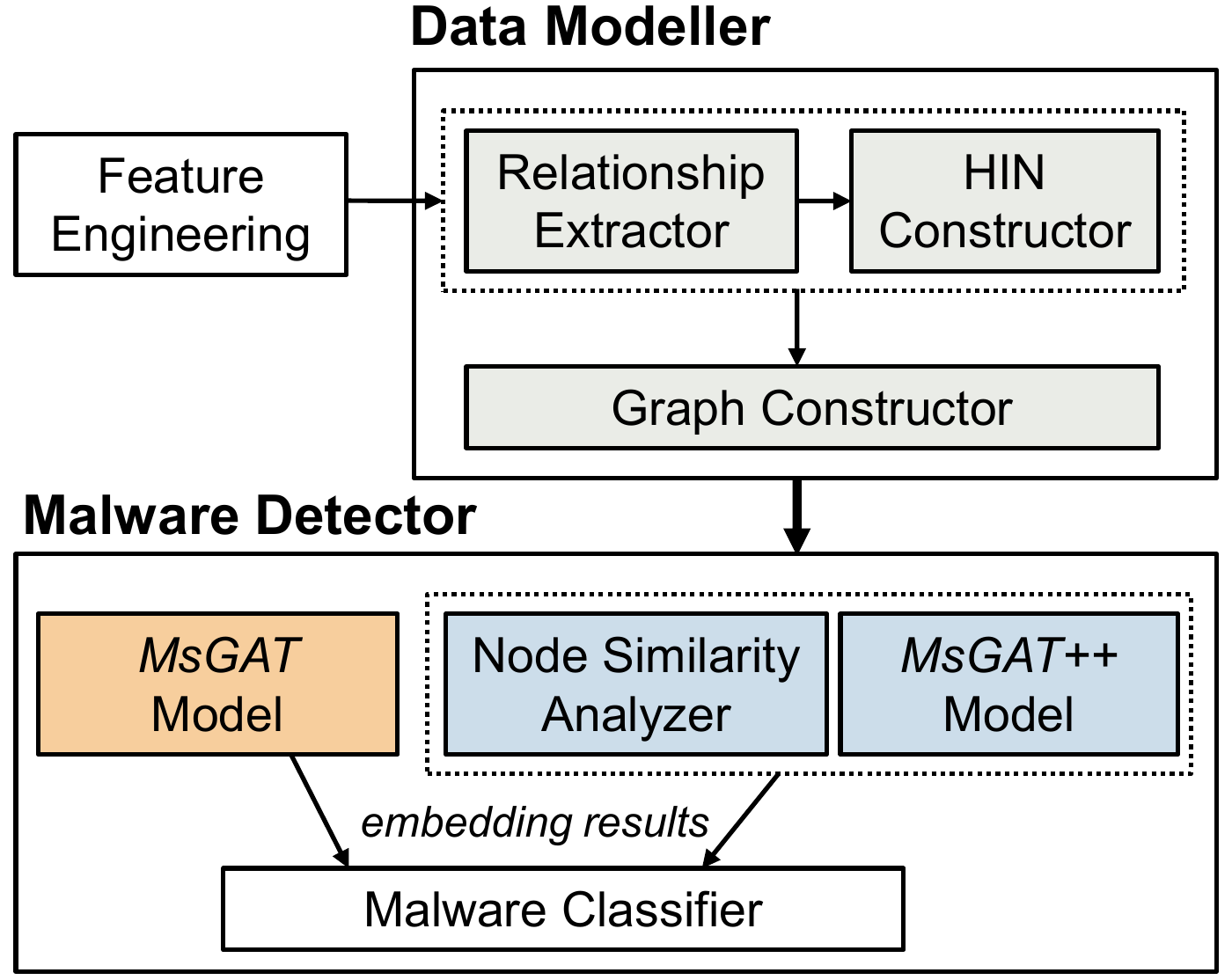}}
\vspace{-1.65mm}
\caption{\projtitle architecture overview}
\vspace{-1.2mm}
\label{fig:sysarch}
\end{figure}

\vspace{-1mm}
\subsection{Our Approach of \projtitle}
\label{sec:background:overview}
\vspace{-0.4mm}

\mypara{Key idea.} 
We consider this problem as a semi-supervised learning based on graph embedding. The first innovation of our approach, as a departure from prior work, is to encode the information as a structured heterogeneous information network (\HIN)~\cite{sun2011pathsim}\cite{peng2021lime} wherein nodes depict entities and their characteristics. A \HIN is a graph $G = (\mathcal{V},\mathcal{E},\mathcal{A},\mathcal{R})$ with an entity type mapping $\phi: \mathcal{V}\to\mathcal{A}$ and a relationship type mapping $\psi:\mathcal{E}\to\mathcal{R}$, where $\mathcal{V}$ and $\mathcal{E}$ represent node and edge set, respectively. $\mathcal{A}$ and $\mathcal{R}$ denote the type set of nodes and edge, where $|\mathcal{A}|$ + $|\mathcal{R}| \textgreater$ 2. Edges represent the relationships between a pair of entities (e.g., an App \textit{owns} a specific permission, or a permission \textit{belongs to} a permission type). 
Since the detection problem is App entity oriented, it is effective to deduce the information from a self-contained \HIN to homogeneous relational subgraphs that can be directly absorbed by GNN. As the fundamental requirement of graph embedding is to obtain the graph structure, we need to calculate the adjacency matrix from the constructed \HIN -- the best option to reflect the proximity and the node connectivity in the graph. GNN models can be subsequently carried out to learn the numerical embedding for in-sample App nodes. To underpin the continuous embedding learning for out-of-sample nodes, the learning model is desired to make the best use of the embedding result of the existing in-sample App nodes, in an incremental manner.




\mypara{Architecture Overview.} Fig.~\ref{fig:sysarch} depicts \projtitle's architecture, encompassing \textit{Data Modeller} and \textit{Malware Detector} components.
Specifically, \textit{Relationship Extractor} in \textit{Data Modeller} firstly offers an  extraction of Android entities based on feature engineering - massive Android Apps are compiled and investigated. There are seven types of nodes ("App" together with six characteristics) and six types of edges. \textit{\HIN Constructor} then builds up the \HIN by organizing entities and the extracted relationships into nodes and edges of \HIN (\S~\ref{sec:hinconstr:relationship}).
\textit{App Graph Constructor} is responsible for generating homogeneous relational subgraphs from 
\HIN that only contains App entities. This is  enabled by employing meta structures including both meta path~\cite{dong2017metapath2vec} and meta graph~\cite{zhang2018metagraph2vec}
(\S~\ref{sec:hinconstr:appgraph}).

\textit{Malware Detector} then involves two distinct representation learning models to numerically embed in-sample and out-of-sample nodes, respectively. It is in great need of fully exploiting node affinities within a given meta-structure and aggregate the embeddings of the same node under different meta-structures. Specifically, we design separate strategies to learn the embedding: 

$\bullet$ To represent in-sample App nodes, the proposed \modelbasic, a meta-structure enabled GAT solution,
firstly aggregate \textit{intra-meta-structure} attention aggregation mechanism for accumulating the embedding of a target node among its neighbor nodes within the graph pertaining to a certain meta-structure. 
In the second \textit{inter-meta-structure} phase, we further fuse the obtained embedding among different meta-structures so that their semantic meanings can be represented in the final embedding (\S~\ref{sec:models:insample}).

$\bullet$ To efficiently tackle the out-of-sample node embedding, we  generate the embedding, \textit{incrementally}, for a new node through reusing and aggregating the embedding result of selective in-sample App nodes in close proximity to the target node. This requires the model to ascertain the similarity between  existing in-sample App nodes and the target node. Similarly, the embedding is firstly gathered at neighbor node level under a given meta-structure before conducting
the \textit{inter-meta-structure} aggregation (\S~\ref{sec:models:outofsample}).

\textit{Malware Classifier} digests the learned vector embeddings to learn a classification model to determine if a given App is malicious or benign and then validates its effectiveness. General purpose techniques such as Random Forest, Logistic Regression, SVM, etc. can be adopted as the classifier implementation. We select the training set from in-sample Apps to train our classifier, whilst using the testing set from in-sample Apps and all out-of-sampling Apps to test the models.

\begin{figure*}[t]
\centerline{\includegraphics[width=0.95\textwidth]{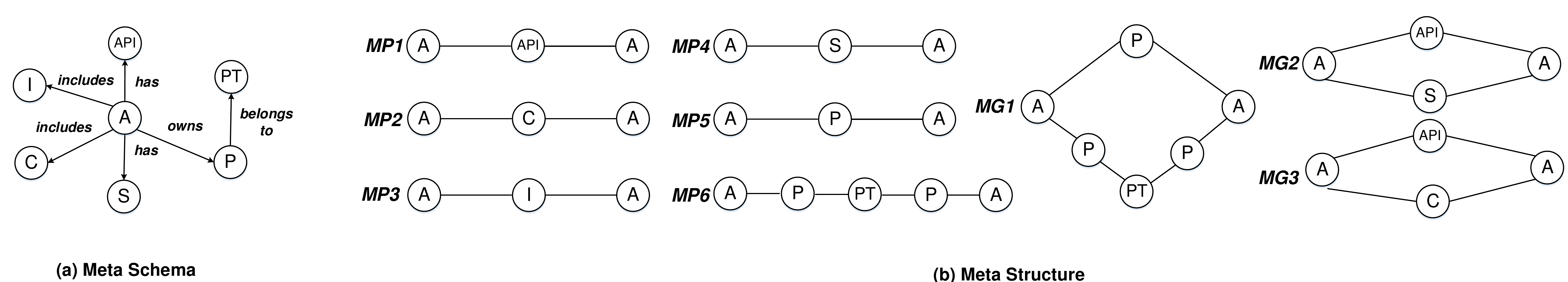}}
\vspace{-0.8em}
\vspace{-1mm}
\caption{(a) Meta-schema and (b) meta-structure}
\label{fig:metafig}
\end{figure*}

\begin{figure}[t]
\centerline{\includegraphics[width=0.48\textwidth]{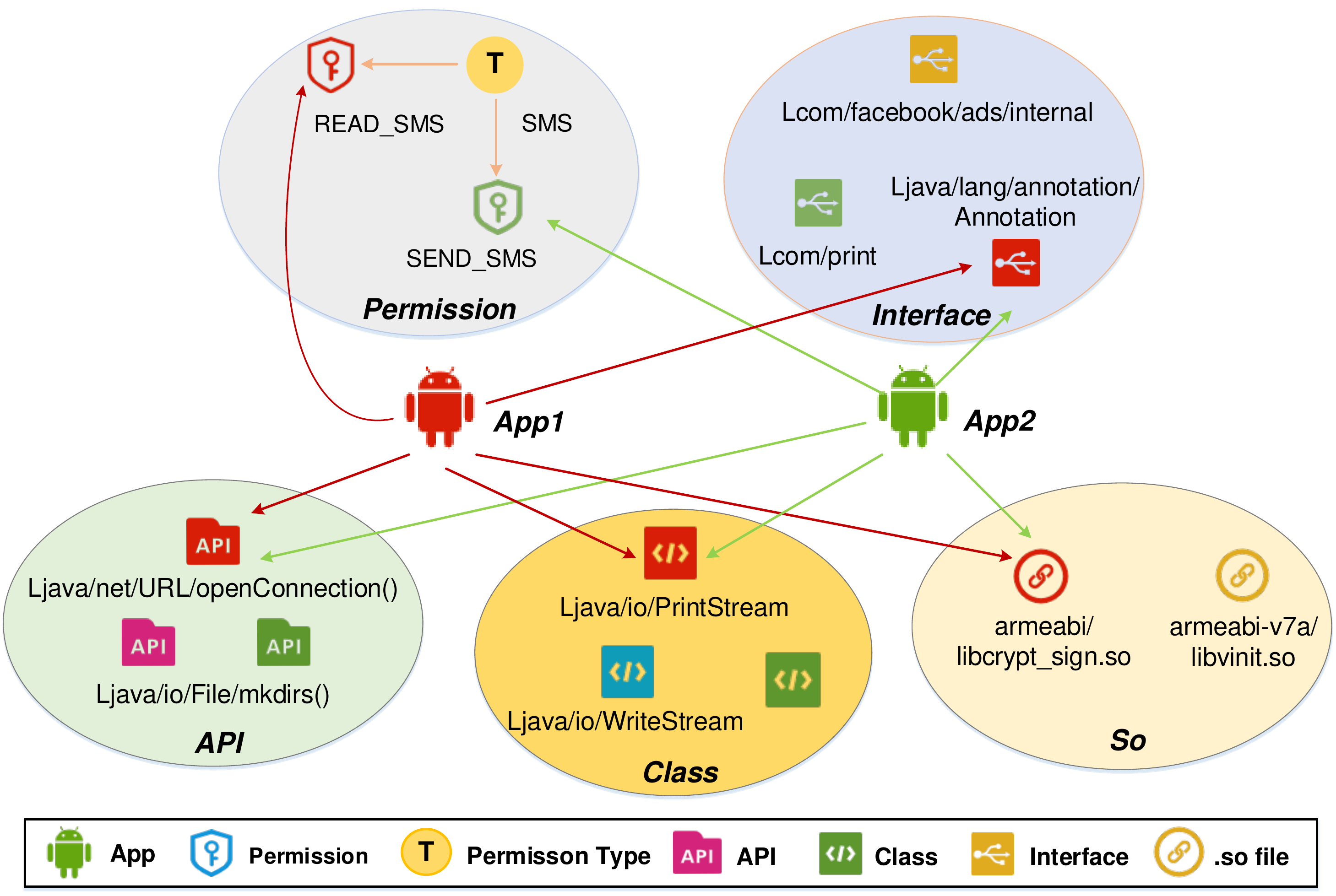}}
\vspace{-1.2mm}
\caption{An example of Android HIN that contains two Android Apps.}
\vspace{-1.2mm}
\label{figmodel}
\end{figure}

\vspace{-0.2mm}
\section{\HIN Based Data Modelling}
\label{sec:hinconstr}

\vspace{-0.1mm}
\subsection{Feature Engineering}
\label{sec:hinconstr:fe}
An Android application needs to be packaged in APK (Android application package) format and installed on Android system. An APK file contains code files, the configuration AndroidManifest.xml file, the signature and verification information, the lib (the directory containing platform-dependent compiled codes) and other resource files. 
To better analyze Android Apps, reverse tools (e.g., APKTool \footnote{https://ibotpeaches.github.io/Apktool}) are widely leveraged to decompile the APK files so that the \texttt{.dex} source file can be decompiled into a \texttt{.smali} file. To describe key characteristics of an App, we extracted the following six types of entities:

$\bullet$  \textbf{Permission (P):} The permission determines specific operations that an App can perform. For example, only Apps with \texttt{READ\_SMS} permission can access user's email information. 

$\bullet$ \textbf{Permission Type (PT):} The permission type \footnote{https://developer.android.google.cn/guide/topics/permissions} describes the category of a given permission. Table~\ref{Table-per} outlines the permission types and representative permissions.

$\bullet$ \textbf{Class (C):} Class is an abstract module in Android codes, where APIs and variables can be directly accesses. \projtitle uses the class name in \texttt{.smali} codes to represent a class.

$\bullet$ \textbf{API:} Application Programming Interface (API) provisions the callable function in Android development environment.  

$\bullet$ \textbf{Interface (I):} The interface refers to an abstract data structure in Java. We extract the name from \texttt{.samli} files.

$\bullet$ \textbf{.so file (S):} \texttt{.so} file is Android's dynamic link library, which can be extracted from the decompiled lib folder.

\begin{table}[t]
	\centering
	\caption{Categories of Representative Permissions}
	\label{Table-per}
    \renewcommand\arraystretch{1.4}
    \vspace{-1mm}
    \footnotesize
    \scalebox{0.94}{
	\begin{tabular}{p{20pt}<{\centering}p{208pt}<{\centering}}
		\toprule
		\textbf{Type}&\textbf{Representative Permissions}\\
		\hline		
       NORMAl&ACCESS_NETWORK_STATE, ACCESS_WIFI_STATE\\
	   CONTACTS&WRITE_CONTACTS, GET_ACCOUNTS\\
       PHONE&READ_CALL_LOG, READ_PHONE_STATE,\\
       CALENDAR&READ_CALENDAR, WRITE_CALENDAR\\
	   LOCATION&ACCESS_FINE_LOCATION, ACCESS_COARSE_LOCATION\\
       STORAGE&READ_EXTERNAL_STORAGE, WRITE_EXTERNAL_STORAGE\\
       SMS&READ_SMS, RECEIVE_MMS, RECEIVE_SMS\\
		\toprule
	\end{tabular}
	}
\end{table}

Following this methodology, we downloaded over 200, 000 APKs from open repositories and after de-duplication and decompilation, 181,235 APKs are finally filtered and extracted. 63,902 entities are then selected according to \cite{li2018significant}. This provisions abundant data sources for establishing the \HIN and mining intrinsic semantics. 

\vspace{-1.5mm}
\subsection{Constructing \HIN}
\vspace{-0.8mm}
\label{sec:hinconstr:relationship}

\mypara{Extracting entity relationships into a \HIN}. Meta-schema is a meta-level template that defines the relationship and type constraints of nodes and edges in the \HIN. As shown in Fig.~\ref{fig:metafig}(a), we figure out a meta-schema that can encode necessary relationships between Android entities.
Based on the domain knowledge, we elaborately examine the following inherent semantic relationships:

$\bullet$ \textbf{[R1] App-API} indicates an App \textit{has} a specific API.  Using the relationship between App and API is effective to dig out and represent the link between two Apps~\cite{3hou2017hindroid}.
	
$\bullet$ \textbf{[R2] App-Permission} specifies an App \textit{owns} a specific permission. Apps with permissions such as \texttt{READ\_SMS}, \texttt{SEND\_SMS}, \texttt{WRITE\_SMS} are strongly correlative~\cite{li2018significant}. If \texttt{SEND\_SMS} is shared between $App_1$ and $App_2$ and \texttt{READ\_SMS} is shared between $App_2$ and $App_3$, an implicit association between $App_1$ and $App_3$ is highly likely to manifest. 
    
$\bullet$ \textbf{[R3] Permission-PermissionType} describes the permission \textit{belongs to} a specific permission type. Normally, permissions can be categorized into different types \footnote{https://developer.android.google.cn/guide/topics/permissions}.
    
$\bullet$ \textbf{[R4] App-Class} means
    the App includes a specific class in the external SDK. A malware tends to generate instances by using classes in a vicious SDK \footnote{https://research.checkpoint.com/2019/simbad-a-rogue-adware-campaign-on-google-play}. 
    
$\bullet$ \textbf{[R5] App-Interface} indicates the App \textit{includes} the specific interface in the external SDK. 

$\bullet$ \textbf{[R6] App-.so} denotes the App \textit{has} a specific \texttt{.so} file. \cite{fan2018gotcha} demonstrates the effectiveness of associating dynamic link libraries with software in Windows system.

Fig.~\ref{figmodel} depicts a \HIN  that contains two Apps and their semantic relationships. For instance, App$_1$ has API \texttt{Ljava/net/URL/openConnection}. Both App$_1$ and App$_2$ own the Class \texttt{Ljava/io/PrintStream}".
The permission \texttt{READ\_SMS} belongs to the permission type \texttt{SMS}", etc. 

\mypara{Storing entity relationships}. We use a relation matrix to store each relationship individually. For instance, 
we generate an matrix $\mathbb{A}$ where the  element $\mathbb{A}_{i,j}$ denotes if App$_i$ contains API$_j$. Intuitively, the transpose of a matrix depicts the backward relationship, e.g., API$_j$ belongs to App$_i$. As summarized in Table \ref{tab:mat}, six matrices are used to represent and store the relationships \textbf{[R1]} to \textbf{[R6]}.
Nevertheless, it is necessary to obtain the connectivity between two Apps if there are sophisticated semantic links, i.e., higher-order relationships. 

\begin{table}[t]
	\centering
	\footnotesize
	\caption{Descriptions of relation matrices.}
	\vspace{-0.8em}
	\label{tab:mat}
    \renewcommand\arraystretch{1.5}
    \scalebox{0.95}{
	\begin{tabular}{p{21pt}<{\centering}p{16pt}<{\centering}p{195pt}}\toprule
		Relation& Matrix& Description\\
		\hline		
        \textbf{R1}&$\mathbb{A}$ & if App $i$ \textit{contains} the API $j$, $a_{i,j}$ is 1;
        otherwise 0.\\
        
	    \textbf{R2}&$\mathbb{P}$ & if App $i$ \textit{has} the permission $j$, $\mathbb{P}_{i,j}$ is 1;
        otherwise 0.\\
        
        \textbf{R3}&$\mathbb{T}$ & if the type of permission $i$ is $j$, $\mathbb{T}_{i,j}$ is 1;
        otherwise  0.\\
        
        \textbf{R4}&$\mathbb{C}$ & if App $i$ owns the Class $j$, $\mathbb{C}_{i,j}$ is 1;
        otherwise 0.\\
        
	    \textbf{R5}&$\mathbb{I}$ & if App $i$ uses the interface $j$, $\mathbb{I}_{i,j}$ is 1;        otherwise 0\\
        \textbf{R6}&$\mathbb{S}$ & if App $i$ calls the so file $j$, $\mathbb{S}_{i,j}$ is 1;
        otherwise 0.\\
		\toprule
	\end{tabular}
	}
\end{table}


\vspace{-0.3mm}
\subsection{Constructing App Graph from \HIN}
\vspace{-0.5mm}
\label{sec:hinconstr:appgraph}

To form a homogeneous graph that only contains App nodes, the key step is to incorporate the relationship between App entity and other entities into the combined connectivity between Apps. To ascertain the hidden higher-order semantic, we mainly calculate Apps' proximity via exploiting a meta-path or meta-graph within a given \HIN and then obtain the node adjacency matrix for the graph. In other words, given a meta structure, the \HIN can be converted to an exclusive homogeneous graph in which each node has meta-structure specific neighbor nodes.

In fact, a \textit{meta-path} connects a pair of nodes with a semantically meaningful relationship. We  enrich the meta-structures further to involve the \textit{meta-graph} -- in the form of directed acyclic graph (DAG) -- that can be used as an extended template to capture arbitrary but meaningful combination of existing relationships between a pair of nodes. In effect,
a meta structure provides a filter view to extract a homogeneous node graph, wherein all nodes satisfy particular complicated semantics.
Arguably, depending upon different meta structures, nodes will be organized distinctly within different graphs. To some extent, each graph can be regarded as a sub-graph of the holistic \HIN under a certain view -- each sub-graph satisfies the semantic constraints given by the meta-structure.

\mypara{Meta structures.} We leverage domain knowledge from system security expertise to elaborately pick up meta structures for covering the inherent relationships. We first combine all possible meaningful semantic meta-structures, and then carefully select those meta-structures with sufficient precision through numerous experiments. The detailed procedure is discussed in \S\ref{sec:exp:microbenchmark}. As shown in Fig.~\ref{fig:metafig}(b), we eventually present  six meta paths and three meta graphs that can effectively outline the structural semantics and capture rich relationships between two Android Apps in the \HIN. For example, \texttt{A-P-A} describes the relationship where two Apps have the same permission ($\mathcal{MP}_5$) and \texttt{A-P-PT-P-A} indicates two Apps co-own the same type of permission ($\mathcal{MP}_6$). $\mathcal{MG}_2$ simultaneously combines \texttt{A-API-A} with \texttt{A-S-A}. Accordingly, the semantic constraints will be tightened, i.e., the selected nodes have to satisfy all pre-defined constraints.  
Nevertheless, models \cite{yun2019graph,hu2020heterogeneous} without the manual design of original meta structures could also be applied into our scheme.

\mypara{Homogeneous App graph for each meta structure.} Performing a sequence of matrix operations over the modeled relationship matrices, we can precisely calculate the adjacency of nodes within a graph. For a given meta-path $\mathcal{MP}$, $(A_1, \dots, A_{n})$, the adjacency matrix can be calculated by
\begin{align}
\label{eq:adjmat4MP}
\footnotesize
\Psi^{\mathcal{MP}} = R_{A_1A_2} \cdot R_{A_2A_3} \dots \cdot R_{A_{n-1}A_{n}}, 
\end{align}
where $R_{A_jA_{j+1}}$ is the relation matrix between entity $A_j$ and $A_{j+1}$ (one instance of \textbf{[R1]} to \textbf{[R6]} in Table~\ref{tab:mat}). 
For example, the adjacency matrix for the graph under $\mathcal{MP}_1$ \texttt{A-API-A} is $\Psi^{\mathcal{MP}_1} = \mathbb{A}\cdot \mathbb{A}^T$.  $\Psi_{i,j} \textgreater 0$ indicates App$_i$ and App$_j$ are associated with each other, i.e., they are neighbors based on the meta-path $\mathcal{MP}_1$. Specifically, the value represents the count of meta-path instances, i.e., the number of pathways, between node $i$ and $j$. 
Likewise, for a given meta-graph
$\mathcal{MG}$, a combination of several meta-paths, i.e., $(\mathcal{MP}_1, \dots, \mathcal{MP}_{m})$, the node adjacency matrix is: 
\begin{align}
\label{eq:adjmat4MG} 
\footnotesize
\Psi^{\mathcal{MG}} = \Psi^{\mathcal{MP}_1} \odot \dots \odot \Psi^{\mathcal{MP}_m}, 
\end{align}
where $\odot$ is the operation of \textit{Hadamard Product}. For instance, $\mathcal{MG}_2$, the adjacency matrix can be calculated by  $\Psi^{\mathcal{MG}_2} = (\mathbb{A}\cdot \mathbb{A}^T) \odot (\mathbb{S}\cdot \mathbb{S}^T)$. By conducting graph modelling for each meta structure, the original \HIN is converted to multiple App homogeneous graphs, each of which pertains to an adjacency matrix. Given $K$ meta-structures, we have a collection of $K$ adjacency matrices, i.e., \{$\Psi^{\mathcal{M}_1}$, ... , $\Psi^{\mathcal{M}_K}$\}.
\begin{table}[t]
	\centering
	\footnotesize
	\caption{Symbol Notations}
    \vspace{-1.6mm}
	\label{tab:symbols}
    \renewcommand\arraystretch{1.5}
    \scalebox{0.88}{
	\begin{tabular}{p{56pt}<{\centering}p{195pt}}
	\toprule
	\textbf{Symbol} & \textbf{Definition}\\ \hline		
    $\mathcal{M}_k$, $\mathcal{MP}$, $\mathcal{MG}$ & $k$th meta-structure,  a meta-path or meta-graph \\  \hline
    $R_{A_iA_j}$ & Relation matrix between two entities in the \HIN\\  \hline
    $Sim_{\mathcal{M}_k}(v_i, v_j)$ & The similarity value between node $v_{i}$ and node $v_{j}$ under meta-structure $\mathcal{M}_k$ \\  \hline		
    $\mathbb{X}_{\mathcal{M}_k}$ & Similarity matrix under meta-structure $\mathcal{M}_k$ \\ \hline
    $\Psi^{\mathcal{M}_k}$ & Adjacency matrix under ${\mathcal{M}_k}$ that can depicts node connectivity in a homo graph\\ \hline	
    $\widehat{\Psi}^{\mathcal{M}_k}$ & incremental segment of the adjacency matrix, connecting in-sample nodes to new nodes \\ \hline 
    $\Phi^{\mathcal{M}_k}$ & Embedding matrix under ${\mathcal{M}_k}$; each single row $\Phi^{\mathcal{M}_k}_i$ represents the vector embedding for $i$th node \\ \hline 
    $\widehat{\Phi}^{\mathcal{M}_k}$ & Embedding matrix under ${\mathcal{M}_k}$ for new nodes \\  
	\toprule
	\end{tabular}
	}
\end{table}

\begin{figure}[t]
\centerline{\includegraphics[width=0.49\textwidth]{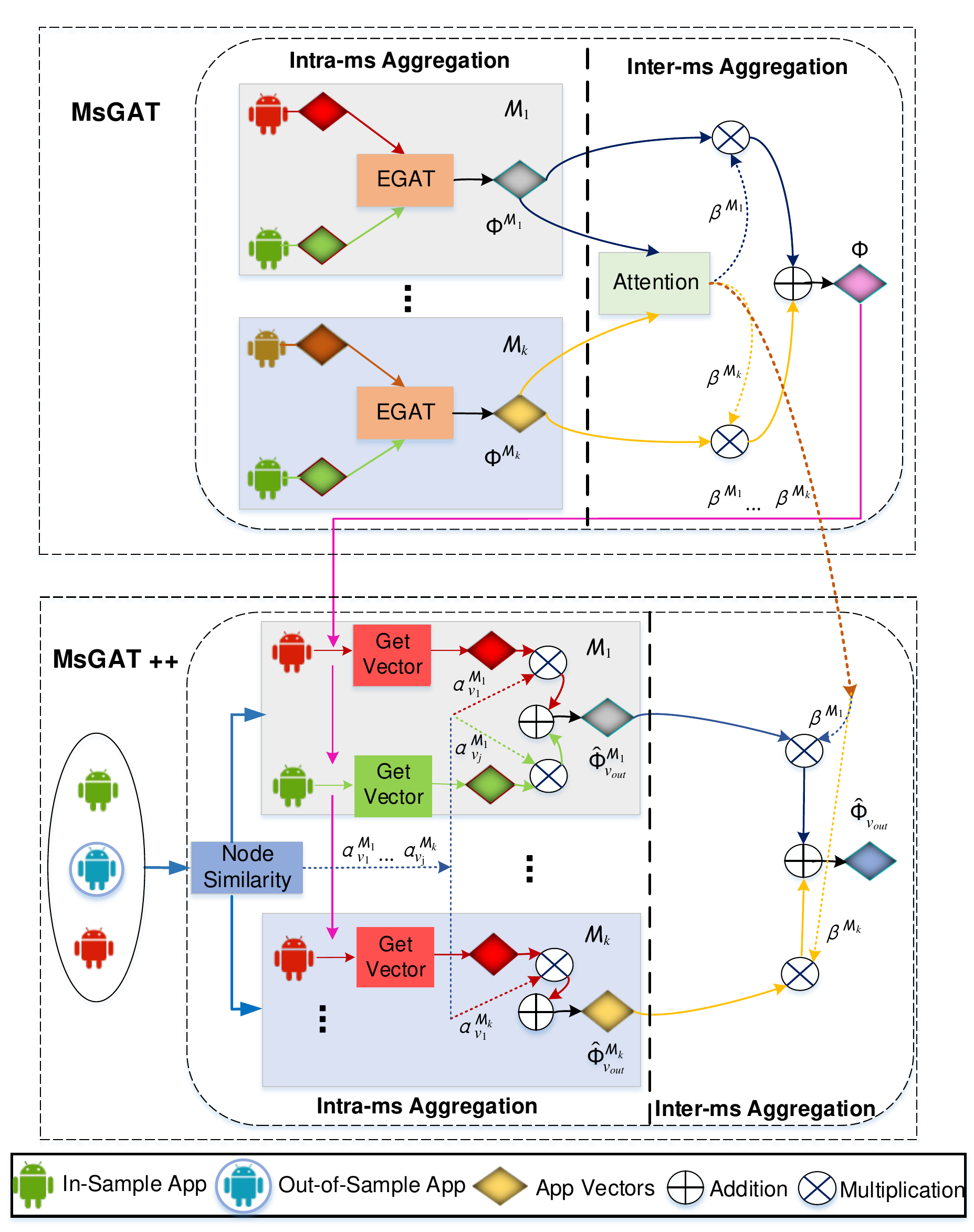}}
\vspace{-1.2mm}
\caption{\modelbasic and \modelplus models for node embedding.}
\vspace{-1mm}
\label{fig:models}
\end{figure}

\vspace{-0.1mm}
\section{Node Embedding Models}
\label{sec:models}

\subsection{\modelbasic: In-Sample Node Embedding} \label{sec:models:insample}
\vspace{-0.4mm}

We introduce a series of innovative Graph Attention Network (GAT) optimizations enhanced by meta-structures -- we employ the attention mechanism ~\cite{vaswani2017attention} among neighbor nodes within a given meta-structure (\textit{intra-ms}) and coordinate the attention among different meta structures (\textit{inter-ms}). 
Fig.~\ref{fig:models} depicts the flowchart of our models and important notations used in the models are outlined in Table~\ref{tab:symbols}.

\mypara{Intra-ms aggregation.} Intra-ms aggregation learns how a node pay different attention to its neighbor nodes in a graph pertaining to a meta-structure. Formally, it aggregates the neighbors' representation vectors with weights considering the feature information of entities and the edge information between entities.
To do so, we initially encode the vector of each in-sample App in the form of one-hot and concatenate them into a matrix $H$.  $H_{i\cdot}$, the $i$th row of $H$, represents the embedding vector of $i$th App node. Thereafter, we design an edge weight aware GAT model (EGAT) to combine $H$ and the adjacency matrix pertaining to a given meta-structure $\mathcal{M}_k$.
To implement the EGAT model, feature information and edge weight information are fully utilized to aggregate features from neighbors. More specifically, we firstly 
construct the adjacency matrix $\Psi^{\mathcal{M'}_k}$ with a normalization operation:
\begin{align}
\label{eqmatrix}
\small
   \Psi^{\mathcal{M'}_k} = Normalize(H \cdot H^{T} \odot \Psi^{\mathcal{M}_k}),
\end{align}
and elements in $\Psi^{\mathcal{M'}_k}$ that are lower than a pre-defined threshold $\tau$ ($\tau$ is set to be 0.1 in our model) will be set zero. 
Thereafter, we update the  $\Phi^{\mathcal{M}_k}$ with GAT model~\cite{velivckovic2017graph}:
\begin{align}
\label{gat}
\footnotesize
   \Phi^{\mathcal{M}_k} = GAT(H; \Psi^{\mathcal{M'}_k}).
\end{align}
Eventually, the low dimensional vector embedding for all in-sample App nodes, in a form of matrix $\Phi^{\mathcal{M}_k}$ with a collection of row vectors, can be obtained in this stage.

We then repeatedly calculate the vector matrix for all pre-defined meta-structures, and obtain a collection of embedding vectors, i.e., [$\Phi^{\mathcal{M}_1}, \dots ,  \Phi^{\mathcal{M}_K}$], where $K$ is the totality of meta-structures. 
Concretely, the embedding matrix $\Phi^{\mathcal{M}_k}$ is of shape  $L \times D$, where $L$ denotes the number of in-sample Apps in the \HIN and $D$ denotes the dimension of each App vector. As a result, the embedding of App$_i$ node can be identified as the $i$th row, i.e., $\Phi^{\mathcal{M}_k}_{i\cdot}$. 

\mypara{Inter-ms aggregation.} Since each
meta structure provisions an individual semantic view, we propose an \textit{inter-ms} attention aggregation to integrate embedding [$\Phi^{\mathcal{M}_1}, \dots ,  \Phi^{\mathcal{M}_K}$] under different semantics and thus enhance the quality of node embedding.  
Specifically, we exploit a multi-layer perceptron (MLP) procedure for learning the weight $\beta^{\mathcal{M}_k}$ of each meta-structure $\mathcal{M}_k$ in the fusion: 
\begin{align}
\label{eq3}
\footnotesize
    (\beta^{\mathcal{M}_1},\dots, \beta^{\mathcal{M}_K}) = \operatorname{ softmax}(\operatorname{NN}(\Phi^{\mathcal{M}_1}),\dots, \operatorname{NN}(\Phi^{\mathcal{M}_K})),
\end{align}
where $\operatorname{NN}$ is a native Neural Network that maps a given matrix to a numerical value.
Consequently, the final embedding for all in-sample App nodes can be obtained through adding up the weighted representation matrices:
\begin{align}
\label{eq4}
\footnotesize
    \Phi = \sum_{k=1}^K \beta^{\mathcal{M}_k} \cdot \Phi^{\mathcal{M}_k}.
\end{align}
we then pass $\Phi$ on to another Neural Network so that the loss function between the Neural Network's outputs and ground-true labels can be calibrated via iterative back-propagation.

\subsection{\modelplus: Incremental  Embedding}
\label{sec:models:outofsample}

To best embed unknown Apps not included in the training procedure, we present \modelplus, an increment  learning mechanism for utilizing the in-sample embedding already learned from  \modelbasic to rapidly represent those out-of-sample Apps. To make clear, we use $v_{out}$ to generally stand for any out-of-sample node out of the \HIN.  

\mypara{Exploring node similarity.} Pinpointing the underlying connections between new nodes and existing nodes in the \HIN plays a pivotal role in providing rapid numerical representation and cost-effective malware detection. To do so, it is imperative to calculate and accumulate the similarity between $v_{out}$ and existing nodes.  
Following similar methodology presented in \cite{gao2020hincti}, the node similarity between node $v_i$ and node $v_j$ under a given meta path is defined as: 
\begin{align}
\label{eq-1}
\footnotesize
Sim^{\mathcal{MP}}(v_i, v_j) = \frac{2*\Psi^{\mathcal{MP}}_{ij}}{\Psi^{\mathcal{MP}}_{ii} + \Psi^{\mathcal{MP}}_{jj}},
\end{align}
where $\Psi^{\mathcal{MP}}_{ij}$ implies the number of meta structures between two connected nodes and thus a higher similarity indicates a tighter association between these two nodes. Accordingly, the node similarity between node $v_i$ and node $v_j$ under a meta graph $\mathcal{MG}$ is:
\begin{align}
\label{eq0} 
\footnotesize
Sim^{\mathcal{MG}}(v_i, v_j) = Sim^{\mathcal{MP}_1}(v_i, v_j) \odot  ...  \odot Sim^{\mathcal{MP}_m}(v_i, v_j).
\end{align}

\mypara{Incremental aggregation for   embedding learning}. The initial task is to catch the incremental relationships and construct the graph information. Within a given meta-structure, we aim to only update an adjacency matrix that quantifies the connectivity between the out-of-sample nodes and existing in-sample App nodes. This should be done in an \textit{incremental} manner to reduce the training cost. In practice, we first repeat the steps aforementioned in \S~\ref{sec:hinconstr:relationship} to calculate all relation matrices in Table~\ref{tab:mat} merely for out-of-sample App nodes. Secondly, we concatenate the relation matrices of new App nodes and those of existing App nodes to form an incremental segment of the node adjacency $\widehat{\Psi}^{\mathcal{M}_k}$ -- a pathway from an in-sample App node to a new node.
Take $\mathcal{MP}_1$ as an example; we first obtain the relation matrix $\mathbb{A}_{out}$ for all new nodes  and then generate the matrix by
$ \widehat{\Psi}^{\mathcal{M}_1} = \mathbb{A}_{in} \cdot \mathbb{A}^T_{out}$. 
This design ensures the incremental adjacency matrix $\widehat{\Psi}^{\mathcal{M}_k}$ can function independently from the established adjacency matrix ${\Psi}^{\mathcal{M}_k}$ whilst they together serve as the holistic abstract of connectivity among all nodes. 

We propose \modelplus to entitle numerical embedding to new nodes whilst calibrating existing node's representation. Similar to 
\modelbasic, the model consists of two steps: \textit{intra-ms} and \textit{inter-ms} aggregation. 
Given a semantic meta-structure $\mathcal{M}_k$, we substitute $\widehat{\Psi}^{\mathcal{M}_k}$
into Eq.~\ref{eq-1} or Eq.~\ref{eq0} to calculate $Sim^{\mathcal{M}_k}(v_j, v_{out})$, the similarity between a new node $v_{out}$ and any in-sample App node $v_j$. Repeating this for all out-of-sampling nodes and all in-sample App nodes forms a similarity matrix  $\mathbb{X}^{\mathcal{M}_k}$ where a larger value inherently indicates a closer proximity between two nodes. Accordingly, we can obtain a collection of similarity matrix for all meta-structures \{$\mathbb{X}^{\mathcal{M}_1},\dots,  \mathbb{X}^{\mathcal{M}_K}$\}.

\begin{algorithm}[t]
    \small
	\caption{Incremental embedding algorithm in \modelplus}
	\label{alg:msgatplus}
	\begin{algorithmic}[1]
		\REQUIRE An out-of-sample App $v_{out}$ \\
		\ENSURE $v_{out}$'s vector embedding $\widehat{\Phi}_{v_{out}}$ and the updated embedding matrix $\Phi$ for existing in-sample App nodes \\
		\FOR {$ k \in \{1, ... , K\}$}
        \STATE \textit{// select $\sigma$ in-sample App nodes with the highest similarity}
        \STATE $\{v_{n1}, \dots, v_{n\sigma}\}$ $\gets$ {\tt DescendSort}($\mathbb{X}^{\mathcal{M}_k}$).\texttt{topK}($\sigma$)
        
		\STATE \textit{// Calculate the weights} 
		
		\STATE \{$\alpha_{v_1}^{\mathcal{M}_k}, \dots, \alpha_{v_\sigma}^{\mathcal{M}_k}$\}  $\gets$ Eq.\ref{eq:plus_weight}
    	
    	\STATE \textit{// Calculate the embedding of $v_{out}$ under $\mathcal{M}_k$}  
    	
    	\STATE $\widehat{\Phi}_{v_{out}}^{\mathcal{M}_k} \gets$ Eq.\ref{eq:plus_aggr}.
		\ENDFOR
		
		\STATE \textit{// Embedding fusion from all meta structures}
		
		\STATE $\widehat{\Phi}_{v_{out}}  \gets$ Eq.~\ref{eq:plus_final}
		
		\RETURN $\widehat{\Phi}_{v_{out}}$, $\Phi$
	\end{algorithmic}
\end{algorithm}

Arguably, to better represent the new node in a numerical vector, we should fully aggregate existing embedding results of existing nodes in closely proximity to the new node. To this end, 
we select top-$\sigma$ in-sample App nodes ($v_{n1}, \dots, v_{n\sigma})$, based on the similarity matrix $\mathbb{X}^{\mathcal{M}_k}$, and aggregate their vectors for 
the embedding of the new node: 
\begin{align}
\label{eq:plus_aggr}
\footnotesize
    \widehat{\Phi}_{v_{out}}^{\mathcal{M}_k} = \sum_{s=1}^\sigma \alpha_{v_{ns}}^{\mathcal{M}_k} \cdot \Phi^{\mathcal{M}_k}_{v_{ns}},
\end{align}
where $\alpha_{v_j}^{\mathcal{M}_k}$ denotes the weight of the node $v_j$ ($v_j \in (v_{n1}, \dots, v_{n\sigma})$) under $\mathcal{M}_k$ and $\widehat{\Phi}$ implies the incremental embedding information for the out-of-sample node exclusively. The weight can be easily calculated by: 
\begin{align}
\label{eq:plus_weight}
\footnotesize
    \alpha_{v_j}^{\mathcal{M}_k} = \frac{Sim^{\mathcal{M}_k}(v_{out}, v_{ns})}{\sum_{s=1}^\sigma Sim^{\mathcal{M}_k}(v_{out}, v_{ns})}.
\end{align}
Eventually, we re-calibrate the embedding by conducting \textit{inter-ms} aggregation over $K$ individual representations under all meta-structures:
\begin{align}\label{eq:plus_final}
\footnotesize
    \widehat{\Phi}_{v_{out}} = \sum_{k=1}^K \beta^{\mathcal{M}_k} \cdot \widehat{\Phi}_{v_{out}}^{\mathcal{M}_k},
\end{align}
where $\beta^{\mathcal{M}_k}$ can be obtained from Eq.~\ref{eq3} (In fact, to improve the performance of our model, we need to fine-tune these weights). Alg.~\ref{alg:msgatplus} outlines the whole procedure of our rapid incremental embedding learning in the malware detection. 

\mypara{Time complexity.} Alg.~\ref{alg:msgatplus} demonstrates a simple but efficient approach with an acceptable complexity. The overall complexity is $\mathcal{O}(KLNlogN)$ where $K$ and $L$ are the number of meta-structures and the number of out-of-sample Apps, respectively while $N$ represents the number of in-sample Apps.


\section{Experiment Setup}
\label{sec:exp_setup}


\subsection{Methodology}

\mypara{Environment.} \projtitle is evaluated on a 16-node GPU cluster, where each node has a 64-core Intel Xeon CPU E5-2680 v4@2.40GHz with 512GB RAM and 8 NVIDIA Tesla P100 GPUs, Ubuntu 20.04 LTS with Linux kernel v.5.4.0. \projtitle depends upon tensorflow-gpu v1.12.0 and scikit-learn v0.21.3. ApkTool and aapt.exe are used for parsing Apps.

\mypara{Datasets.} 
According to the aforementioned discussion of feature engineering in \S\ref{sec:hinconstr:fe}, we overall decompiled 181,235 APKs (i.e., 80,860 malicious Apps and 100,375 benign Apps) from 2013 to 2019. with the help of AndroZoo\footnote{https://androzoo.uni.lu}, benign Apps are primarily collected from GooglePlay store while malicious Apps are obtained from VirusShare and CICAndMal.
To validate the compatibility, both forward and backward, of the proposed model in \projtitle, we train our model based on Apps released in 2017 (amid the seven time span), and then utilize it to detect Apps published from 2013 to 2019. 

Specifically, we extracted 14,000 benign and 9,865 malicious Apps released in 2017, as in-sample Apps, to construct the \HIN and train the detection model. For generating the out-of-sample sample data, we collected 7 malware subsets (\textit{v2013} to \textit{v2019}), each of which contains  roughly 10,000 samples, from VirusShare over consecutive seven years, together with another 2 subsets from CICAndMal, including 242 scarewares/adwares samples in 2017 (\textit{c2017}) and 253 samples in 2019 (\textit{c2019}). Meanwhile, we extracted benign Apps to match the same number of benign Apps in each subset above.   

\mypara{Methodology and Metrics.} The experiments are three-fold: we firstly evaluate the effectiveness of \projtitle against traditional feature-based  ML approaches and numerous baselines in terms of in-sample and out-of-sample scenarios (\S\ref{sec:exp:effectivenss}). Afterwards, we demonstrate the efficiency of \projtitle by comparing the training time consumption with other approaches (\S\ref{sec:exp:efficiency}). 
We further conduct several micro-benchmarkings, including an ablation analysis of performance gains, an evaluation of meta-structure's importance and the impact of the sampled neighbor number on detection precision (\S\ref{sec:exp:microbenchmark}). 

We use metrics $Precision$, $Recall$, $FP$-$Rate$, $F1$ and $Accurate$ to measure the effectiveness (see Table~\ref{Metrics}), and use time consumption to measure the efficiency. The execution time includes the process of generating embedding vectors and detecting Apps whilst excluding the process of extracting Apps relation matrix. We use 5-fold cross validation and calculate the average accuracy to provide an assurance of unbiased and accurate evaluation.

\begin{table}[t]
	\centering
	\vspace{-0.8em}
	\caption{Descriptions of evaluation metrics.}
	\label{Metrics}
    \renewcommand\arraystretch{1.4}
	\vspace{-0.8em}
    \scalebox{0.88}{
	\begin{tabular}{p{43pt}<{\centering}p{209pt}<{\centering}}\toprule
		Metrics&Description\\
		\hline		
        $TP$&The number of malicious Apps that are correctly identified\\
	    $TN$&The number of benign Apps that are correctly identified\\
        $FP$&The number of benign Apps that are mistakenly identified\\
	    $TN$&The number of malicious Apps that are mistakenly identified\\
        $Precision$&$TP/(TP+FP)$\\
	    $Recall$&$TP/(TP+FN)$\\
	    $FP$-$Rate$&$FP/(FP+TN)$\\
        $F1$&$2*Precision$*$Recall/(Precision+Recall)$\\
	    $Acc$&$(TP+FN)/(TP+TN+FP+FN)$\\
		\toprule
	\end{tabular}
	}
\end{table}

\vspace{-1mm}
\subsection{Baselines} 

To evaluate the performance of \modelbasic in \projtitle, the baselines encompasses generic models and specific models used by some well-known malware detection systems. 

\mypara{Generic models.} We firstly implement the following generic models as comparative approaches:
	
		\noindent $\bullet$ \textbf{Node2Vec} \cite{grover2016node2vec} is a typical  model generalized from DeepWalk \cite{perozzi2014deepwalk} based on homogeneous graph network.

		\noindent $\bullet$ \textbf{GCN} \cite{kipf2016semi} is a semi-supervised homogeneous graph convolutional network model that retains feature information and structure information of the graph nodes.

        \noindent $\bullet$ \textbf{RS-GCN} represents the approach to converting the \HIN into homogeneous graphs, applying native GCN to each graph and reporting the best performance among different graphs.

    	\noindent $\bullet$ \textbf{GAT} \cite{velivckovic2017graph} is a semi-supervised homogeneous graph model that utilizes attention mechanism for aggregating neighborhood information of graph nodes.
    
        \noindent $\bullet$ \textbf{RS-GAT} denotes the approach to converting the \HIN into homogeneous graphs based on rich semantic meta-structures, applying native GAT to each homogeneous graph and reporting the best performance among different graphs.
    	
    	\noindent$\bullet$ \textbf{Metapath2Vec} \cite{dong2017metapath2vec} is a heterogeneous graph representation learning model that leverages  meta-path based random walk to find neighborhood and uses skip-gram with negative sampling to learn node vectors.
    
        \noindent$\bullet$ \textbf{Metagraph2Vec}~\cite{zhang2018metagraph2vec} is an alternative model to Metapath2Vec; both meta paths and meta graphs are applied to  the random walk.
    
        \noindent $\bullet$ \textbf{HAN}~\cite{wang2019heterogeneous} is a heterogeneous graph representation learning model that utilizes predefined meta paths and hierarchical attentions for node vector embedding.

For Node2Vec, GCN and GAT, we treat all the nodes in \HIN as the same type to obtain the homogeneous graph. Since all these models are towards static graphs, we compare the capability of out-of-sample detection between \modelplus and three generic strategies that can be easily adopted in any comparative models:

		\noindent $\bullet$ \textbf{Neighbor averaging (NA)} directly averages the vector embedding of the in-sample neighbors pertaining to a given new App as the targeted embedding.

        \noindent $\bullet$ \textbf{Sampled neighbor averaging (SNA)} further filters the neighbor range by sampling a fixed number of in-sample neighbors based on the sorted node similarity and simply averaging their embedding as the targeted embedding. 
		
	    \noindent $\bullet$ \textbf{Re-running (RR)} primarily merges the out-of-sample Apps with in-sample Apps and rebuilds the entire \HIN and the malware detection model.

\mypara{Specific models deriving from specialized systems.} Secondly, we compare our models in \projtitle against the following models used by the existing malware detection systems:

    \noindent $\bullet$ \textbf{Drebin}~\cite{arp2014drebin} is a framework that inspects a given App by extracting a wide range of features sets from the \texttt{manifest} and \texttt{dex} code and adopts the SVM model in the classifier.
    
    \noindent $\bullet$ \textbf{DroidEvolver}~\cite{xu2019droidevolver} 
    is a self-evolving detection system to maintain and rely on a model pool of different detection models that are initialized with a set of labeled Apps using various online learning algorithms. It is worth noting that we do not directly compare against MamaDroid~\cite{mariconti2016mamadroid}, because it has been demonstrated less effective than DroidEvolver. 
    
    \noindent $\bullet$ \textbf{HinDroid}~\cite{3hou2017hindroid} constructs a heterogeneous graph with entities such as App and API and and the rich in-between relationships. It aggregates information from different semantic meta-paths and uses multi-kernel learning to calculate the representations of Apps. 
    
    \noindent $\bullet$ \textbf{MatchGNet}~\cite{mgnet} is a graph-based malware detection model that regards each software as a heterogeneous graph and learns its representation. It determines the threat of an unknown software primarily through matching the graph representation of the unknown software and that of benign software.

    \noindent \textbf{Aidroid}~\cite{ye2018aidroid} is among the first attempts to tackle out-of-sample malware representations with heterogeneous graph model and CNN network. Following the detailed description in the paper, we utilize one-hop and two-hop neighbors to best function its model performance.

\mypara{Model parameters}. For Node2Vec and Metapath2Vec, we set the number of walks per node, the max walk length, and the window size to be 10, 100, 8, respectively.
For GCN, GAT and HAN, we set up the parameters suggested by their original papers. For the fairness of comparison, each model will be trained 200 times. The length of embedding vectors delivered by these models are set to  be 128.

\begin{table}[t]
	\centering
	\caption{The F1 Value and Accuracy of In-sample Apps Detection.}
	\label{In-sample}
	\vspace{-2mm}
    \renewcommand\arraystretch{1.3}
    \scalebox{0.94}{
	\begin{tabular}{p{14pt}<{\centering}p{65pt}<{\centering}p{22pt}<{\centering}p{22pt}<{\centering}p{22pt}<{\centering}p{22pt}<{\centering}}\toprule
		Metrics&Approaches&20\%&40\%&60\%&80\%\\
		\midrule		
        $\multirow{14}{*}{\rotatebox{90}{$F1$}}$
	    &Node2Vec&0.8355&0.8378&0.8542&0.8601\\
        &GCN&0.8653&0.8677&0.8721&0.8763\\
        &GAT&0.8435&0.8633&0.8752&0.8801\\
	    &Metapath2Vec&0.9231&0.9321&0.9328&0.9395\\
	    &RS-GCN&0.9212&0.9510&0.9515&0.9560\\
        &RS-GAT&0.9507&0.9631&0.9653&0.9664\\
        &HAN&0.9511&0.9617&0.9671&0.9705\\
	    &Metagraph2Vec&0.9750&0.9766&0.9764&0.9771\\
	    &SVM (Drebin)&0.9312&0.9387&0.9446&0.9477\\ 
	    &DroidEvolver&0.9412&0.9517&0.9566&0.9605\\
	    &HinDroid&0.9643&0.9669&0.9684&0.9746\\
	    &MatchGNet&0.9395&0.9511&0.9604&0.9753\\
	    &Aidroid&0.9321&0.9399&0.9414&0.9455\\
        &\modelbasic (\projtitle) &\textbf{0.9857}&\textbf{0.9859}&\textbf{0.9871}&\textbf{0.9878}\\
        \midrule
        $\multirow{14}{*}{\rotatebox{90}{$Acc$}}$
	    &Node2Vec&0.8254&0.8388&0.8405&0.8593\\
        &GCN&0.8558&0.8663&0.8630&0.8692\\
        &GAT&0.8461&0.8645&0.8758&0.8833\\
	    &Metapath2Vec&0.9259&0.9321&0.9335&0.9388\\
	    &RS-GCN&0.9199&0.9494&0.9527&0.9544\\
        &RS-GAT&0.9486&0.9620&0.9652&0.9664\\
        &HAN&0.9521&0.9657&0.9675&0.9699\\
	    &Metagraph2Vec&0.9686&0.9698&0.9748&0.9762\\
	    &SVM (Drebin)&0.9295&0.9356&0.9407&0.9455\\
	    &DroidEvolver&0.9329&0.9506&0.9557&0.9623\\
	    &HinDroid&0.9688&0.9698&0.9722&0.9764\\
	    &MatchGNet&0.9302&0.9508&0.9536&0.9689\\
	    &Aidroid&0.9227&0.9356&0.9367&0.9437\\
        &\modelbasic (\projtitle) &\textbf{0.9843}&\textbf{0.9855}&\textbf{0.9867}&\textbf{0.9854}\\
        \toprule
	\end{tabular}
	}
\end{table}

\begin{table}[t]
	\centering
	\caption{The F-P Rate of In-sample Apps Detection.}
	\label{In-sample-FP}
	\vspace{-2mm}
    \renewcommand\arraystretch{1.3}
    \scalebox{0.94}{
	\begin{tabular}{p{14pt}<{\centering}p{65pt}<{\centering}p{22pt}<{\centering}p{22pt}<{\centering}p{22pt}<{\centering}p{22pt}<{\centering}}\toprule
		Metrics&Approaches&20\%&40\%&60\%&80\%\\
        \midrule
        $\multirow{14}{*}{\rotatebox{90}{$FP-Rate$}}$
	    &Node2Vec&0.0425&0.0393&0.0388&0.0342\\
        &GCN&0.0350&0.0323&0.0333&0.0318\\
        &GAT&0.0343&0.0334&0.0299&0.0268\\
	    &Metapath2Vec&0.0177&0.0175&0.0169&0.0165\\
	    &RS-GCN&0.0184&0.0118&0.0109&0.0107\\
        &RS-GAT&0.0115&0.0088&0.0079&0.0075\\
        &HAN&0.0108&0.0098&0.0085&0.0087\\
	    &Metagraph2Vec&0.0071&0.0068&0.0059&0.0057\\
	    &SVM (Drebin)&0.0163&0.0155&0.0135&0.0139\\
	    &DroidEvolver&0.0154&0.0116&0.0101&0.0108\\
	    &HinDroid&0.0075&0.0078&0.0071&0.0068\\
	    &MatchGNet&0.0193&0.0129&0.0122&0.0081\\
	    &Aidroid&0.0184&0.0171&0.0150&0.0139\\
        &\modelbasic (\projtitle) &\textbf{0.0038}&\textbf{0.0034}&\textbf{0.0032}&\textbf{0.0035}\\
        \toprule
	\end{tabular}
	}
\end{table}

\begin{table*}[t]
	\centering
	\caption{The F1 Value of out-of-sample Apps Detection.}
	\vspace{-2mm}
	\label{table:out_of_sample_fone}
    \renewcommand\arraystretch{1.4}
    \footnotesize
    \scalebox{0.97}{
	\begin{tabular}{p{19pt}<{\centering}p{54pt}<{\centering}p{67pt}<{\centering}p{23pt}<{\centering}p{23pt}<{\centering}p{23pt}<{\centering}p{23pt}<{\centering}p{23pt}<{\centering}p{23pt}<{\centering}p{23pt}<{\centering}p{23pt}<{\centering}p{23pt}<{\centering}}\toprule
		Metrics&In-sample Approaches& Out-of-sample Approaches &v2013&v2014&v2015&v2016&v2017&v2018&v2019&c2017&c2019\\
		\midrule		
        $\multirow{30}{*}{\rotatebox{90}{$F1$}}$
        &$\multirow{3}{*}{Node2Vec}$&NA&0.5888&0.6746&0.6965&0.6740&0.6811&0.6744&0.6680&0.6533&0.6995\\
        &&SNA&0.6541&0.6732&0.6965&0.6935&0.6851&0.6665&0.6685&0.6638&0.6845\\
        &&Rerunning&0.7564&0.8102&0.7956&0.8124&0.8236&0.7549&0.7968&0.7765&0.7945\\

        &\multicolumn{1}{>{\columncolor{mycyan}}c}{GCN}&\multicolumn{1}{>{\columncolor{mycyan}}c}{Rerunning}&\multicolumn{1}{>{\columncolor{mycyan}}c}{0.8637}&\multicolumn{1}{>{\columncolor{mycyan}}c}{0.8705}&\multicolumn{1}{>{\columncolor{mycyan}}c}{0.8459}&\multicolumn{1}{>{\columncolor{mycyan}}c}{0.8496}&\multicolumn{1}{>{\columncolor{mycyan}}c}{0.8697}&\multicolumn{1}{>{\columncolor{mycyan}}c}{0.8743}&\multicolumn{1}{>{\columncolor{mycyan}}c}{0.8637}&\multicolumn{1}{>{\columncolor{mycyan}}c}{0.8567}&\multicolumn{1}{>{\columncolor{mycyan}}c}{0.8537}\\

         &$\multirow{3}{*}{GAT}$&NA&0.7364&0.7423&0.7153&0.7155&0.7545&0.6225&0.7203&0.6352&0.6442\\
        &&SNA&0.7433&0.7521&0.7056&0.6962&0.6842&0.7121&0.6831&0.6720&0.6318\\
        &&Rerunning&0.8242&0.8448&0.8531&0.8474&0.8731&0.8595&0.8457&0.8511&0.8476\\

        &\multicolumn{1}{>{\columncolor{mycyan}}c}{}&\multicolumn{1}{>{\columncolor{mycyan}}c}{NA}&\multicolumn{1}{>{\columncolor{mycyan}}c}{0.7414}&\multicolumn{1}{>{\columncolor{mycyan}}c}{0.8424}&\multicolumn{1}{>{\columncolor{mycyan}}c}{0.7835}&\multicolumn{1}{>{\columncolor{mycyan}}c}{0.7784}&\multicolumn{1}{>{\columncolor{mycyan}}c}{0.7537}&\multicolumn{1}{>{\columncolor{mycyan}}c}{0.8243}&\multicolumn{1}{>{\columncolor{mycyan}}c}{0.8473}&\multicolumn{1}{>{\columncolor{mycyan}}c}{0.8160}&\multicolumn{1}{>{\columncolor{mycyan}}c}{0.8183}\\
        &\multicolumn{1}{>{\columncolor{mycyan}}c}{Metapath2Vec}&\multicolumn{1}{>{\columncolor{mycyan}}c}{SNA}&\multicolumn{1}{>{\columncolor{mycyan}}c}{0.7564}&\multicolumn{1}{>{\columncolor{mycyan}}c}{0.8531}&\multicolumn{1}{>{\columncolor{mycyan}}c}{0.7765}&\multicolumn{1}{>{\columncolor{mycyan}}c}{0.7496}&\multicolumn{1}{>{\columncolor{mycyan}}c}{0.7365}&\multicolumn{1}{>{\columncolor{mycyan}}c}{0.8359}&\multicolumn{1}{>{\columncolor{mycyan}}c}{0.8363}&\multicolumn{1}{>{\columncolor{mycyan}}c}{0.8242}&\multicolumn{1}{>{\columncolor{mycyan}}c}{0.8156}\\
        &\multicolumn{1}{>{\columncolor{mycyan}}c}{}&\multicolumn{1}{>{\columncolor{mycyan}}c}{Rerunning}&\multicolumn{1}{>{\columncolor{mycyan}}c}{0.9240}&\multicolumn{1}{>{\columncolor{mycyan}}c}{0.9321}&\multicolumn{1}{>{\columncolor{mycyan}}c}{0.9195}&\multicolumn{1}{>{\columncolor{mycyan}}c}{0.9214}&\multicolumn{1}{>{\columncolor{mycyan}}c}{0.9342}&\multicolumn{1}{>{\columncolor{mycyan}}c}{0.9326}&\multicolumn{1}{>{\columncolor{mycyan}}c}{0.9285}&\multicolumn{1}{>{\columncolor{mycyan}}c}{0.9094}&\multicolumn{1}{>{\columncolor{mycyan}}c}{0.9052}\\

        &$\multirow{4}{*}{HAN}$&NA&0.7455&0.7405&0.6361&0.7433&0.7292&0.7443&0.7245&0.7101&0.7253\\
        &&SNA&0.7593&0.7635&0.7793&0.7723&0.8046&0.7803&0.7566&0.7543&0.7768\\
        &&Rerunning&0.9155&0.9626&0.9678&0.9588&0.9758&0.9522&0.9677&0.9482&0.9574\\
        &&\modelplus&0.8896&0.9611&0.9512&0.9462&0.9466&0.9655&0.9583&0.9358&0.9386\\

        &\multicolumn{1}{>{\columncolor{mycyan}}c}{RS-GCN}&\multicolumn{1}{>{\columncolor{mycyan}}c}{Rerunning}&\multicolumn{1}{>{\columncolor{mycyan}}c}{0.9532}&\multicolumn{1}{>{\columncolor{mycyan}}c}{0.9549}&\multicolumn{1}{>{\columncolor{mycyan}}c}{0.9487}&\multicolumn{1}{>{\columncolor{mycyan}}c}{0.9499}&\multicolumn{1}{>{\columncolor{mycyan}}c}{0.9656}&\multicolumn{1}{>{\columncolor{mycyan}}c}{0.9651}&\multicolumn{1}{>{\columncolor{mycyan}}c}{0.9745}&\multicolumn{1}{>{\columncolor{mycyan}}c}{0.9539}&\multicolumn{1}{>{\columncolor{mycyan}}c}{0.9471}\\

        &$\multirow{3}{*}{RS-GAT}$&NA&0.7564&0.9400&0.8104&0.6755&0.7345&0.6423&0.7520&0.6152&0.5931\\
        &&SNA&0.7564&0.9400&0.8601&0.6744&0.5290&0.7253&0.7323&0.5807&0.7707\\
        &&Rerunning&0.9260&0.9321&0.9428&0.9582&0.9498&0.9392&0.9372&0.9485&0.9593\\

        &\multicolumn{1}{>{\columncolor{mycyan}}c}{}&\multicolumn{1}{>{\columncolor{mycyan}}c}{NA}&\multicolumn{1}{>{\columncolor{mycyan}}c}{0.7658}&\multicolumn{1}{>{\columncolor{mycyan}}c}{0.9763}&\multicolumn{1}{>{\columncolor{mycyan}}c}{0.8041}&\multicolumn{1}{>{\columncolor{mycyan}}c}{0.7955}&\multicolumn{1}{>{\columncolor{mycyan}}c}{0.7693}&\multicolumn{1}{>{\columncolor{mycyan}}c}{0.8665}&\multicolumn{1}{>{\columncolor{mycyan}}c}{0.7614}&\multicolumn{1}{>{\columncolor{mycyan}}c}{0.8267}&\multicolumn{1}{>{\columncolor{mycyan}}c}{0.8084}\\
        &\multicolumn{1}{>{\columncolor{mycyan}}c}{Metagraph2Vec}&\multicolumn{1}{>{\columncolor{mycyan}}c}{SNA}&\multicolumn{1}{>{\columncolor{mycyan}}c}{0.7672}&\multicolumn{1}{>{\columncolor{mycyan}}c}{0.7769}&\multicolumn{1}{>{\columncolor{mycyan}}c}{0.8155}&\multicolumn{1}{>{\columncolor{mycyan}}c}{0.7996}&\multicolumn{1}{>{\columncolor{mycyan}}c}{0.7805}&\multicolumn{1}{>{\columncolor{mycyan}}c}{0.8665}&\multicolumn{1}{>{\columncolor{mycyan}}c}{0.7628}&\multicolumn{1}{>{\columncolor{mycyan}}c}{0.8239}&\multicolumn{1}{>{\columncolor{mycyan}}c}{0.8084}\\
        &\multicolumn{1}{>{\columncolor{mycyan}}c}{}&\multicolumn{1}{>{\columncolor{mycyan}}c}{Rerunning}&\multicolumn{1}{>{\columncolor{mycyan}}c}{0.9533}&\multicolumn{1}{>{\columncolor{mycyan}}c}{0.9688}&\multicolumn{1}{>{\columncolor{mycyan}}c}{0.9255}&\multicolumn{1}{>{\columncolor{mycyan}}c}{0.9382}&\multicolumn{1}{>{\columncolor{mycyan}}c}{0.9201}&\multicolumn{1}{>{\columncolor{mycyan}}c}{0.9667}&\multicolumn{1}{>{\columncolor{mycyan}}c}{0.9718}&\multicolumn{1}{>{\columncolor{mycyan}}c}{0.9234}&\multicolumn{1}{>{\columncolor{mycyan}}c}{0.9040}\\
        
        &$\multirow{1}{*}{Drebin}$ & &0.7442&0.7723&0.7856&0.8277&0.9432&0.7761&0.7891&0.7559&0.7413\\
        
        &\multicolumn{1}{>{\columncolor{mycyan}}c}{DroidEvolver}&\multicolumn{1}{>{\columncolor{mycyan}}c}{}&\multicolumn{1}{>{\columncolor{mycyan}}c}{0.7972}&\multicolumn{1}{>{\columncolor{mycyan}}c}{0.8469}&\multicolumn{1}{>{\columncolor{mycyan}}c}{0.8519}&\multicolumn{1}{>{\columncolor{mycyan}}c}{0.8996}&\multicolumn{1}{>{\columncolor{mycyan}}c}{0.9605}&\multicolumn{1}{>{\columncolor{mycyan}}c}{0.9265}&\multicolumn{1}{>{\columncolor{mycyan}}c}{0.9028}&\multicolumn{1}{>{\columncolor{mycyan}}c}{0.8539}&\multicolumn{1}{>{\columncolor{mycyan}}c}{0.8584}\\
        
        &$\multirow{1}{*}{HinDroid}$ & &0.8946&0.9232&0.9298&0.9277&0.9712&0.9159&0.9466&0.9396&0.9245\\
        
        &\multicolumn{1}{>{\columncolor{mycyan}}c}{MatchGNet}&\multicolumn{1}{>{\columncolor{mycyan}}c}{}&\multicolumn{1}{>{\columncolor{mycyan}}c}{0.8981}&\multicolumn{1}{>{\columncolor{mycyan}}c}{0.8965}&\multicolumn{1}{>{\columncolor{mycyan}}c}{0.9323}&\multicolumn{1}{>{\columncolor{mycyan}}c}{0.8833}&\multicolumn{1}{>{\columncolor{mycyan}}c}{0.9675}&\multicolumn{1}{>{\columncolor{mycyan}}c}{0.9265}&\multicolumn{1}{>{\columncolor{mycyan}}c}{0.9053}&\multicolumn{1}{>{\columncolor{mycyan}}c}{0.9123}&\multicolumn{1}{>{\columncolor{mycyan}}c}{0.9137}\\

        &$\multirow{1}{*}{HGiNE (AiDroid)}$ &HG2Img &0.8842&0.9723&0.9556&0.9272&0.9455&0.8761&0.8991&0.8959&0.9013\\

        &\multicolumn{1}{>{\columncolor{mycyan}}c}{}&\multicolumn{1}{>{\columncolor{mycyan}}c}{NA}&\multicolumn{1}{>{\columncolor{mycyan}}c}{0.7693}&\multicolumn{1}{>{\columncolor{mycyan}}c}{0.7601}&\multicolumn{1}{>{\columncolor{mycyan}}c}{0.6465}&\multicolumn{1}{>{\columncolor{mycyan}}c}{0.7725}&\multicolumn{1}{>{\columncolor{mycyan}}c}{0.7693}&\multicolumn{1}{>{\columncolor{mycyan}}c}{0.7741}&\multicolumn{1}{>{\columncolor{mycyan}}c}{0.7741}&\multicolumn{1}{>{\columncolor{mycyan}}c}{0.7401}&\multicolumn{1}{>{\columncolor{mycyan}}c}{0.7454}\\
        &\multicolumn{1}{>{\columncolor{mycyan}}c}{\modelbasic}&\multicolumn{1}{>{\columncolor{mycyan}}c}{SNA}&\multicolumn{1}{>{\columncolor{mycyan}}c}{0.7795}&\multicolumn{1}{>{\columncolor{mycyan}}c}{0.7845}&\multicolumn{1}{>{\columncolor{mycyan}}c}{0.7996}&\multicolumn{1}{>{\columncolor{mycyan}}c}{0.8058}&\multicolumn{1}{>{\columncolor{mycyan}}c}{0.8241}&\multicolumn{1}{>{\columncolor{mycyan}}c}{0.7955}&\multicolumn{1}{>{\columncolor{mycyan}}c}{0.7832}&\multicolumn{1}{>{\columncolor{mycyan}}c}{0.7791}&\multicolumn{1}{>{\columncolor{mycyan}}c}{0.8071}\\
        &\multicolumn{1}{>{\columncolor{mycyan}}c}{}&\multicolumn{1}{>{\columncolor{mycyan}}c}{Rerunning}&\multicolumn{1}{>{\columncolor{mycyan}}c}{\textbf{0.9569}}&\multicolumn{1}{>{\columncolor{mycyan}}c}{\textbf{0.9824}}&\multicolumn{1}{>{\columncolor{mycyan}}c}{\textbf{0.9876}}&\multicolumn{1}{>{\columncolor{mycyan}}c}{\textbf{0.9720}}&\multicolumn{1}{>{\columncolor{mycyan}}c}{\textbf{0.9769}}&\multicolumn{1}{>{\columncolor{mycyan}}c}{\textbf{0.9808}}&\multicolumn{1}{>{\columncolor{mycyan}}c}{\textbf{0.9805}}&\multicolumn{1}{>{\columncolor{mycyan}}c}{\textbf{0.9621}}&\multicolumn{1}{>{\columncolor{mycyan}}c}{\textbf{0.9693}}\\
        &\multicolumn{1}{>{\columncolor{mycyan}}c}{}&\multicolumn{1}{>{\columncolor{mycyan}}c}{\modelplus}&\multicolumn{1}{>{\columncolor{mycyan}}c}{0.9007}&\multicolumn{1}{>{\columncolor{mycyan}}c}{0.9804}&\multicolumn{1}{>{\columncolor{mycyan}}c}{0.9736}&\multicolumn{1}{>{\columncolor{mycyan}}c}{0.9687}&\multicolumn{1}{>{\columncolor{mycyan}}c}{0.9695}&\multicolumn{1}{>{\columncolor{mycyan}}c}{0.9665}&\multicolumn{1}{>{\columncolor{mycyan}}c}{0.9658}&\multicolumn{1}{>{\columncolor{mycyan}}c}{0.9461}&\multicolumn{1}{>{\columncolor{mycyan}}c}{0.9393}\\

        \toprule
	\end{tabular}
	}
    \end{table*}

    \begin{table*}[t]
	\centering
	\caption{The False Positive rate of out-of-sample Apps Detection.}
	\vspace{-2mm}
	\label{table:out_of_sample_FPR}
    \renewcommand\arraystretch{1.4}
    \footnotesize
    \scalebox{0.97}{
	\begin{tabular}{p{19pt}<{\centering}p{54pt}<{\centering}p{67pt}<{\centering}p{23pt}<{\centering}p{23pt}<{\centering}p{23pt}<{\centering}p{23pt}<{\centering}p{23pt}<{\centering}p{23pt}<{\centering}p{23pt}<{\centering}p{23pt}<{\centering}p{23pt}<{\centering}}\toprule
		Metrics&In-sample Approaches& Out-of-sample Approaches &v2013&v2014&v2015&v2016&v2017&v2018&v2019&c2017&c2019\\
		\midrule		
        $\multirow{30}{*}{\rotatebox{90}{$FP-Rate$}}$
        &$\multirow{3}{*}{Node2Vec}$&NA&0.1052&0.0846&0.0819&0.0782&0.0776&0.0846&0.0763&0.0971&0.0819\\
        &&SNA&0.0968&0.0831&0.0758&0.0811&0.0862&0.0883&0.0852&0.0806&0.0789\\
        &&Rerunning&0.0682&0.0531&0.0576&0.0534&0.0508&0.0698&0.0579&0.0643&0.0569\\

        &\multicolumn{1}{>{\columncolor{mycyan}}c}{GCN}&\multicolumn{1}{>{\columncolor{mycyan}}c}{Rerunning}&\multicolumn{1}{>{\columncolor{mycyan}}c}{0.0377}&\multicolumn{1}{>{\columncolor{mycyan}}c}{0.0359}&\multicolumn{1}{>{\columncolor{mycyan}}c}{0.0428}&\multicolumn{1}{>{\columncolor{mycyan}}c}{0.0412}&\multicolumn{1}{>{\columncolor{mycyan}}c}{0.0366}&\multicolumn{1}{>{\columncolor{mycyan}}c}{0.0356}&\multicolumn{1}{>{\columncolor{mycyan}}c}{0.0374}&\multicolumn{1}{>{\columncolor{mycyan}}c}{0.0394}&\multicolumn{1}{>{\columncolor{mycyan}}c}{0.0406}\\

         &$\multirow{3}{*}{GAT}$&NA&0.0711&0.0708&0.0754&0.0736&0.0648&0.0981&0.0727&0.0963&0.0911\\
        &&SNA&0.0675&0.0655&0.0779&0.0804&0.0836&0.0754&0.0830&0.0859&0.0966\\
        &&Rerunning&0.0461&0.0408&0.0387&0.0403&0.0334&0.0370&0.0406&0.0394&0.0403\\

        &\multicolumn{1}{>{\columncolor{mycyan}}c}{}&\multicolumn{1}{>{\columncolor{mycyan}}c}{NA}&\multicolumn{1}{>{\columncolor{mycyan}}c}{0.0690}&\multicolumn{1}{>{\columncolor{mycyan}}c}{0.0419}&\multicolumn{1}{>{\columncolor{mycyan}}c}{0.0575}&\multicolumn{1}{>{\columncolor{mycyan}}c}{0.0593}&\multicolumn{1}{>{\columncolor{mycyan}}c}{0.0655}&\multicolumn{1}{>{\columncolor{mycyan}}c}{0.0460}&\multicolumn{1}{>{\columncolor{mycyan}}c}{0.0398}&\multicolumn{1}{>{\columncolor{mycyan}}c}{0.0474}&\multicolumn{1}{>{\columncolor{mycyan}}c}{0.0459}\\
        &\multicolumn{1}{>{\columncolor{mycyan}}c}{Metapath2Vec}&\multicolumn{1}{>{\columncolor{mycyan}}c}{SNA}&\multicolumn{1}{>{\columncolor{mycyan}}c}{0.0616}&\multicolumn{1}{>{\columncolor{mycyan}}c}{0.0371}&\multicolumn{1}{>{\columncolor{mycyan}}c}{0.0565}&\multicolumn{1}{>{\columncolor{mycyan}}c}{0.0634}&\multicolumn{1}{>{\columncolor{mycyan}}c}{0.0667}&\multicolumn{1}{>{\columncolor{mycyan}}c}{0.0416}&\multicolumn{1}{>{\columncolor{mycyan}}c}{0.0415}&\multicolumn{1}{>{\columncolor{mycyan}}c}{0.0455}&\multicolumn{1}{>{\columncolor{mycyan}}c}{0.0467}\\
        &\multicolumn{1}{>{\columncolor{mycyan}}c}{}&\multicolumn{1}{>{\columncolor{mycyan}}c}{Rerunning}&\multicolumn{1}{>{\columncolor{mycyan}}c}{0.0192}&\multicolumn{1}{>{\columncolor{mycyan}}c}{0.0173}&\multicolumn{1}{>{\columncolor{mycyan}}c}{0.0205}&\multicolumn{1}{>{\columncolor{mycyan}}c}{0.0201}&\multicolumn{1}{>{\columncolor{mycyan}}c}{0.0167}&\multicolumn{1}{>{\columncolor{mycyan}}c}{0.0171}&\multicolumn{1}{>{\columncolor{mycyan}}c}{0.0182}&\multicolumn{1}{>{\columncolor{mycyan}}c}{0.0230}&\multicolumn{1}{>{\columncolor{mycyan}}c}{0.0241}\\

        &$\multirow{4}{*}{HAN}$&NA&0.0644&0.0657&0.0921&0.0650&0.0686&0.0647&0.0701&0.0737&0.0701\\
        &&SNA&0.0614&0.0603&0.0563&0.0581&0.0496&0.0559&0.7566&0.0625&0.0568\\
        &&Rerunning&0.0215&0.0094&0.0091&0.0104&0.0061&0.0121&0.0081&0.0131&0.0108\\
        &&\modelplus&0.0279&0.0098&0.0123&0.0136&0.0135&0.0087&0.0105&0.0162&0.0165\\

        &\multicolumn{1}{>{\columncolor{mycyan}}c}{RS-GCN}&\multicolumn{1}{>{\columncolor{mycyan}}c}{Rerunning}&\multicolumn{1}{>{\columncolor{mycyan}}c}{0.0119}&\multicolumn{1}{>{\columncolor{mycyan}}c}{0.0115}&\multicolumn{1}{>{\columncolor{mycyan}}c}{0.0131}&\multicolumn{1}{>{\columncolor{mycyan}}c}{0.0127}&\multicolumn{1}{>{\columncolor{mycyan}}c}{0.0087}&\multicolumn{1}{>{\columncolor{mycyan}}c}{0.0088}&\multicolumn{1}{>{\columncolor{mycyan}}c}{0.0065}&\multicolumn{1}{>{\columncolor{mycyan}}c}{0.0117}&\multicolumn{1}{>{\columncolor{mycyan}}c}{0.0134}\\

        &$\multirow{3}{*}{RS-GAT}$&NA&0.0619&0.0153&0.0484&0.0822&0.0672&0.0906&0.0628&0.0975&0.1039\\
        &&SNA&0.0622&0.0153&0.0358&0.0835&0.1203&0.0702&0.0683&0.1071&0.0585\\
        &&Rerunning&0.0189&0.0172&0.0145&0.0106&0.0127&0.0154&0.1586&0.0130&0.0106\\

        &\multicolumn{1}{>{\columncolor{mycyan}}c}{}&\multicolumn{1}{>{\columncolor{mycyan}}c}{NA}&\multicolumn{1}{>{\columncolor{mycyan}}c}{0.0591}&\multicolumn{1}{>{\columncolor{mycyan}}c}{0.0059}&\multicolumn{1}{>{\columncolor{mycyan}}c}{0.0494}&\multicolumn{1}{>{\columncolor{mycyan}}c}{0.0521}&\multicolumn{1}{>{\columncolor{mycyan}}c}{0.0586}&\multicolumn{1}{>{\columncolor{mycyan}}c}{0.0339}&\multicolumn{1}{>{\columncolor{mycyan}}c}{0.0607}&\multicolumn{1}{>{\columncolor{mycyan}}c}{0.0441}&\multicolumn{1}{>{\columncolor{mycyan}}c}{0.0485}\\
        &\multicolumn{1}{>{\columncolor{mycyan}}c}{Metagraph2Vec}&\multicolumn{1}{>{\columncolor{mycyan}}c}{SNA}&\multicolumn{1}{>{\columncolor{mycyan}}c}{0.0591}&\multicolumn{1}{>{\columncolor{mycyan}}c}{0.0565}&\multicolumn{1}{>{\columncolor{mycyan}}c}{0.0467}&\multicolumn{1}{>{\columncolor{mycyan}}c}{0.0507}&\multicolumn{1}{>{\columncolor{mycyan}}c}{0.0556}&\multicolumn{1}{>{\columncolor{mycyan}}c}{0.0338}&\multicolumn{1}{>{\columncolor{mycyan}}c}{0.0599}&\multicolumn{1}{>{\columncolor{mycyan}}c}{0.0444}&\multicolumn{1}{>{\columncolor{mycyan}}c}{0.0483}\\
        &\multicolumn{1}{>{\columncolor{mycyan}}c}{}&\multicolumn{1}{>{\columncolor{mycyan}}c}{Rerunning}&\multicolumn{1}{>{\columncolor{mycyan}}c}{0.0117}&\multicolumn{1}{>{\columncolor{mycyan}}c}{0.0079}&\multicolumn{1}{>{\columncolor{mycyan}}c}{0.0188}&\multicolumn{1}{>{\columncolor{mycyan}}c}{0.0156}&\multicolumn{1}{>{\columncolor{mycyan}}c}{0.0202}&\multicolumn{1}{>{\columncolor{mycyan}}c}{0.0084}&\multicolumn{1}{>{\columncolor{mycyan}}c}{0.0071}&\multicolumn{1}{>{\columncolor{mycyan}}c}{0.0196}&\multicolumn{1}{>{\columncolor{mycyan}}c}{0.0242}\\
        
        &$\multirow{1}{*}{Drebin}$ & &0.0653&0.0583&0.0547&0.0440&0.0145&0.0572&0.0538&0.0623&0.0653\\
        
        &\multicolumn{1}{>{\columncolor{mycyan}}c}{DroidEvolver}&\multicolumn{1}{>{\columncolor{mycyan}}c}{}&\multicolumn{1}{>{\columncolor{mycyan}}c}{0.0517}&\multicolumn{1}{>{\columncolor{mycyan}}c}{0.0391}&\multicolumn{1}{>{\columncolor{mycyan}}c}{0.0376}&\multicolumn{1}{>{\columncolor{mycyan}}c}{0.0255}&\multicolumn{1}{>{\columncolor{mycyan}}c}{0.0101}&\multicolumn{1}{>{\columncolor{mycyan}}c}{0.0187}&\multicolumn{1}{>{\columncolor{mycyan}}c}{0.0248}&\multicolumn{1}{>{\columncolor{mycyan}}c}{0.0372}&\multicolumn{1}{>{\columncolor{mycyan}}c}{0.0365}\\
        
        &$\multirow{1}{*}{HinDroid}$ & &0.0241&0.0177&0.0253&0.0157&0.0061&0.0201&0.0149&0.0153&0.0162\\
        
        &\multicolumn{1}{>{\columncolor{mycyan}}c}{MatchGNet}&\multicolumn{1}{>{\columncolor{mycyan}}c}{}&\multicolumn{1}{>{\columncolor{mycyan}}c}{0.0257}&\multicolumn{1}{>{\columncolor{mycyan}}c}{0.0218}&\multicolumn{1}{>{\columncolor{mycyan}}c}{0.0137}&\multicolumn{1}{>{\columncolor{mycyan}}c}{0.0236}&\multicolumn{1}{>{\columncolor{mycyan}}c}{0.0065}&\multicolumn{1}{>{\columncolor{mycyan}}c}{0.0156}&\multicolumn{1}{>{\columncolor{mycyan}}c}{0.0201}&\multicolumn{1}{>{\columncolor{mycyan}}c}{0.0185}&\multicolumn{1}{>{\columncolor{mycyan}}c}{0.0173}\\

        &$\multirow{1}{*}{HGiNE (AiDroid)}$ &HG2Img &0.0295&0.0071&0.0113&0.0185&0.0139&0.0316&0.0257&0.0265&0.0252\\

        &\multicolumn{1}{>{\columncolor{mycyan}}c}{}&\multicolumn{1}{>{\columncolor{mycyan}}c}{NA}&\multicolumn{1}{>{\columncolor{mycyan}}c}{0.0589}&\multicolumn{1}{>{\columncolor{mycyan}}c}{0.0608}&\multicolumn{1}{>{\columncolor{mycyan}}c}{0.0895}&\multicolumn{1}{>{\columncolor{mycyan}}c}{0.0576}&\multicolumn{1}{>{\columncolor{mycyan}}c}{0.0584}&\multicolumn{1}{>{\columncolor{mycyan}}c}{0.0572}&\multicolumn{1}{>{\columncolor{mycyan}}c}{0.0577}&\multicolumn{1}{>{\columncolor{mycyan}}c}{0.0659}&\multicolumn{1}{>{\columncolor{mycyan}}c}{0.0648}\\
        &\multicolumn{1}{>{\columncolor{mycyan}}c}{\modelbasic}&\multicolumn{1}{>{\columncolor{mycyan}}c}{SNA}&\multicolumn{1}{>{\columncolor{mycyan}}c}{0.0561}&\multicolumn{1}{>{\columncolor{mycyan}}c}{0.0549}&\multicolumn{1}{>{\columncolor{mycyan}}c}{0.0510}&\multicolumn{1}{>{\columncolor{mycyan}}c}{0.0494}&\multicolumn{1}{>{\columncolor{mycyan}}c}{0.0448}&\multicolumn{1}{>{\columncolor{mycyan}}c}{0.0521}&\multicolumn{1}{>{\columncolor{mycyan}}c}{0.0552}&\multicolumn{1}{>{\columncolor{mycyan}}c}{0.0563}&\multicolumn{1}{>{\columncolor{mycyan}}c}{0.0491}\\
        &\multicolumn{1}{>{\columncolor{mycyan}}c}{}&\multicolumn{1}{>{\columncolor{mycyan}}c}{Rerunning}&\multicolumn{1}{>{\columncolor{mycyan}}c}{\textbf{0.0109}}&\multicolumn{1}{>{\columncolor{mycyan}}c}{\textbf{0.0044}}&\multicolumn{1}{>{\columncolor{mycyan}}c}{\textbf{0.0032}}&\multicolumn{1}{>{\columncolor{mycyan}}c}{\textbf{0.0071}}&\multicolumn{1}{>{\columncolor{mycyan}}c}{\textbf{0.0058}}&\multicolumn{1}{>{\columncolor{mycyan}}c}{\textbf{0.0049}}&\multicolumn{1}{>{\columncolor{mycyan}}c}{\textbf{0.0049}}&\multicolumn{1}{>{\columncolor{mycyan}}c}{\textbf{0.0097}}&\multicolumn{1}{>{\columncolor{mycyan}}c}{\textbf{0.0078}}\\
        &\multicolumn{1}{>{\columncolor{mycyan}}c}{}&\multicolumn{1}{>{\columncolor{mycyan}}c}{\modelplus}&\multicolumn{1}{>{\columncolor{mycyan}}c}{0.0232}&\multicolumn{1}{>{\columncolor{mycyan}}c}{0.0049}&\multicolumn{1}{>{\columncolor{mycyan}}c}{0.0067}&\multicolumn{1}{>{\columncolor{mycyan}}c}{0.0079}&\multicolumn{1}{>{\columncolor{mycyan}}c}{0.0077}&\multicolumn{1}{>{\columncolor{mycyan}}c}{0.0085}&\multicolumn{1}{>{\columncolor{mycyan}}c}{0.0086}&\multicolumn{1}{>{\columncolor{mycyan}}c}{0.0136}&\multicolumn{1}{>{\columncolor{mycyan}}c}{0.0154}\\

        \toprule
	\end{tabular}
	}
    \end{table*}
    
 \begin{figure*}[htbp]
\centering
\subfigure[F1 Score] {\label{fig:maF1}
\begin{minipage}[t]{0.5\linewidth}
\centering
\includegraphics[width=0.9\columnwidth]{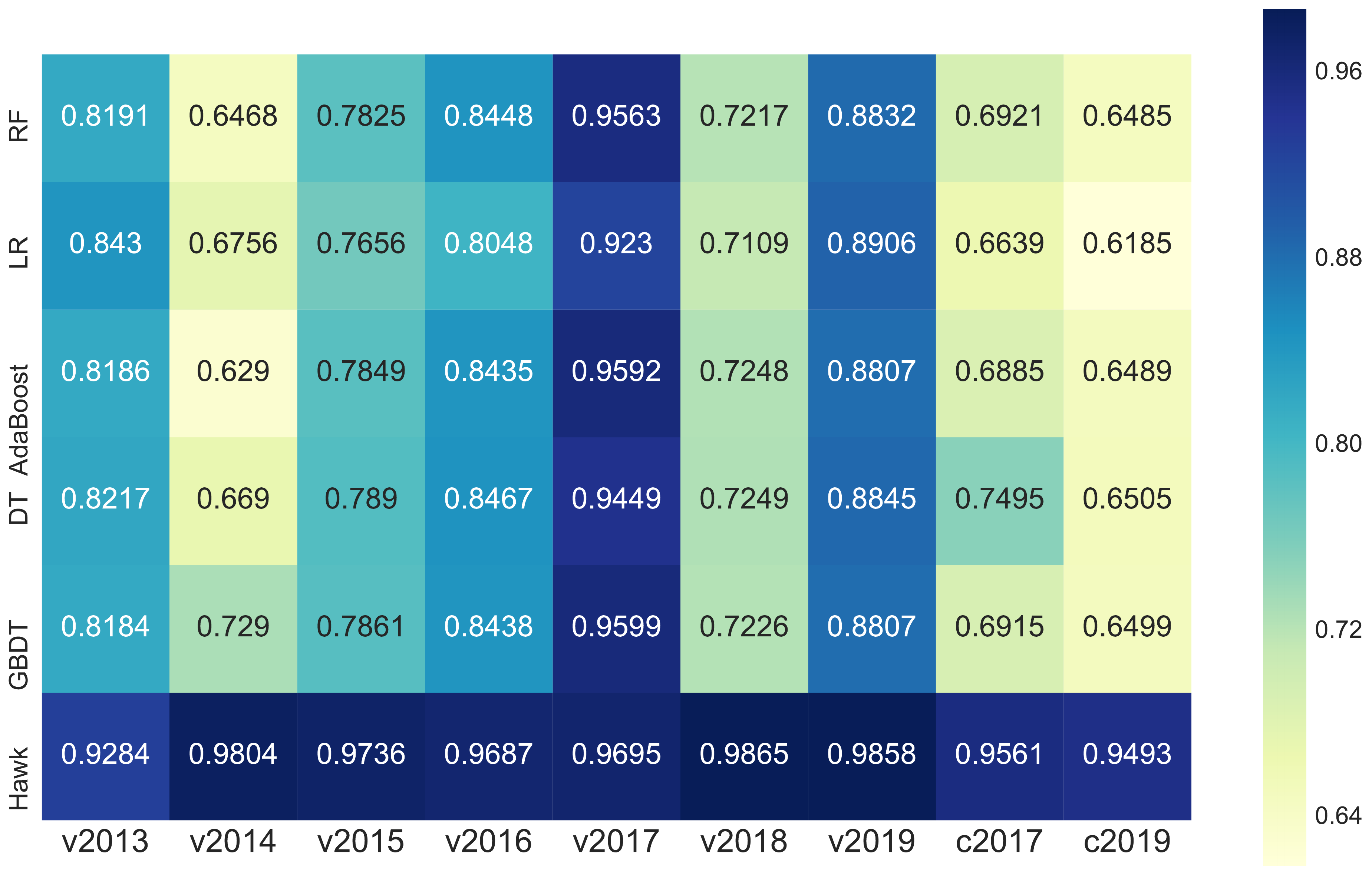}
\end{minipage}%
}%
\subfigure[Acc Score] {\label{fig:maAcc}
\begin{minipage}[t]{0.5\linewidth}
\centering
\includegraphics[width=0.9\columnwidth]{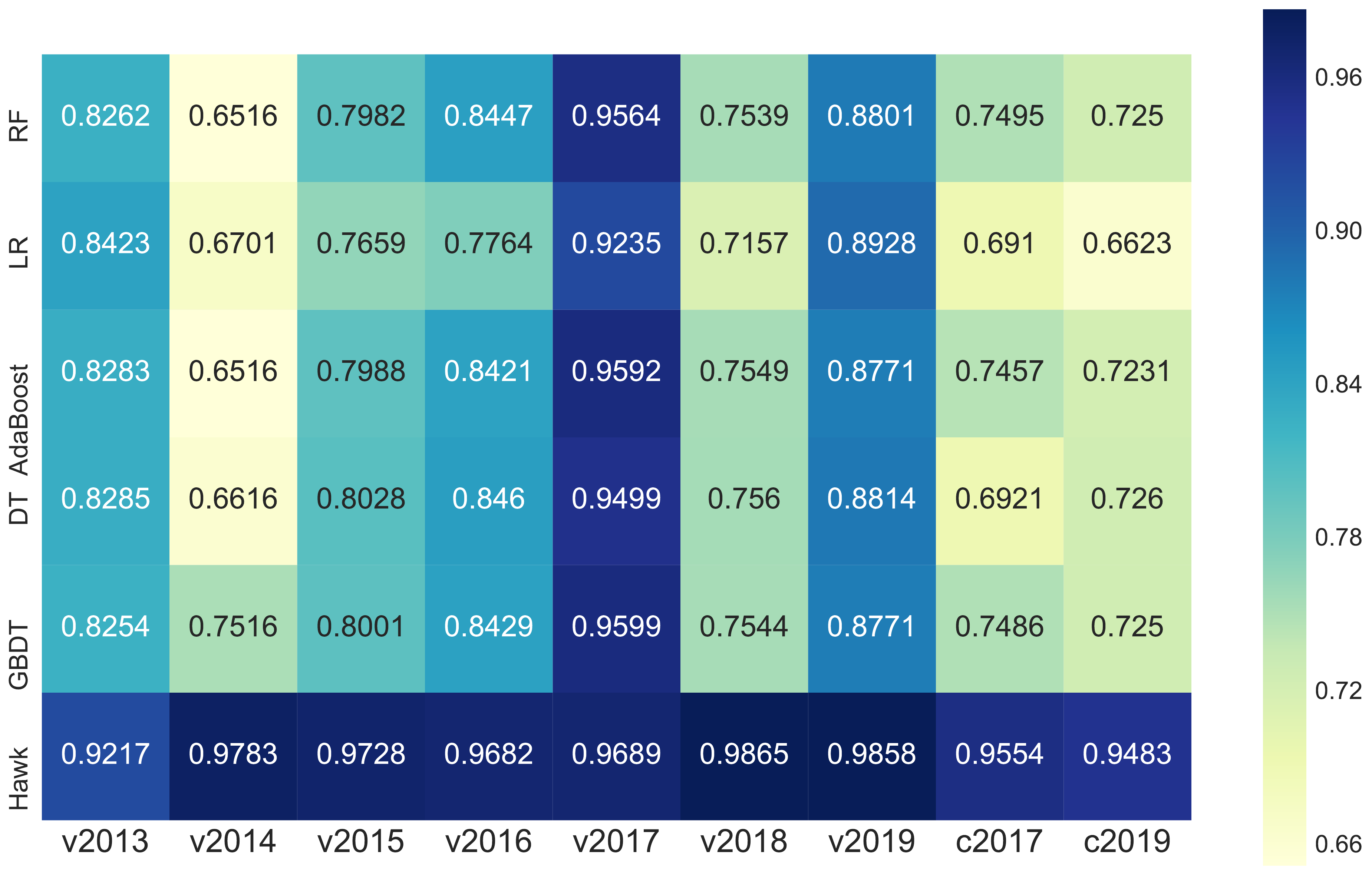}
\end{minipage}%
}%

\centering
\caption{Comparisons with Traditional Machine Learning Methods.}
\label{machine}
\end{figure*}


\section{Experiment Results}
\label{sec:exp}

\subsection{Detection Effectiveness}
\label{sec:exp:effectivenss}

\mypara{In-sample malware detection against DL models}. We choose 20\%, 40\%, 60\%, 80\% of the in-sample Apps to train the Logistic Regression model and the residual for testing. Table~\ref{In-sample} illustrates the F1 and Acc scores of each models. In general, \modelbasic can achieve competitive classification accuracy when compared the popular malware detectors such as Drebin, DroidEvolver, MatchGNet, HinDroid and AiDroid. Compared with F1 and Acc scores, similar observations can be found in Table~\ref{In-sample-FP} when measuring False Positive rate. This is because our graph-based representation learning models can fully integrate the feature information of Apps and the implied semantic information between Apps, which improves the expression ability. In addition, the accuracy of RS-GCN and RS-GAT can be improved by over 5\% compared with native GCN and GAT. Such approaches convert the original \HIN into homogeneous graph and the improvement  derives from preserving the semantic information in the heterogeneous networks through our proposed semantic meta-structures.

It is worth noting that Metagraph2Vec and \modelbasic achieve the highest precision, particularly compared against Metapath2Vec and HAN that only involve meta-paths. The accuracy gain, obviously, stems from introducing meta-graphs that bring rich semantics to mine more complex semantic associations. In addition, \modelbasic outperforms Metagraph2Vec as our models adopt the aggregation mechanisms for both inter-meta-structure and intra-meta-structure, thereby aggregating semantic information from far more comprehensive views.
 
\mypara{Out-of-sample malware detection against DL models}. 
Table~\ref{table:out_of_sample_fone} and Table~\ref{table:out_of_sample_FPR} show the F1 score and False Positive rate, respectively, when we adopt different in-sample models and out-of-sample policies.  Overall, the NA and SNA policies have the lowest detection accuracy under all cases due to the substantial loss of semantic information. Obviously, direct averaging operation ignores the discrepancies  among neighbors thereby reducing the precision of node embedding and the resultant detection effectiveness. 
It is also observable that NA and SNA have very similar precision in almost all cases. This indicates
sampling a certain number of neighbor nodes is able to achieve approximate information in comparison to averaging all neighbor nodes. 

Intuitively, the re-running policy will deliver the best performance of detection over all datasets since all data either new or old will involve in the embedding retraining. Metagraph2Vec, RS-GAT and RS-GCN outperforms Metapath2Vec, GAT and GCN due to the benefit from abundant meta-structures. 
This performance improvement again demonstrates applying abundant semantic meta-structures into embedding models can bring a stronger generalization capacity. 

As shown in Table~\ref{table:out_of_sample_fone}, \modelbasic, together with the rerunning policy, achieves the best detection effectiveness on 2/3 datasets. This can be attributed to the highly rich meta-structures used to include all possible contributions from both intra- and inter- meta-structure aspects. Nevertheless, rerunning has non-negligible overheads particularly in terms of long training time (we will demonstrate the time consumption later). By contrast, \modelplus is proved to be a compromising but competitive solution; the precision of \modelplus is in close proximity to the rerunning baselines over all datasets. 
To demonstrate the generalization, we also implement our \modelplus
mechanism upon the HAN model. Similarly, the incremental learning scheme makes far better improvements when compared against native NA and SNA, only with neglectable margin from the rerunning baseline.  

Hindroid, MatchGNet, HG2Img and Drebin observably deliver unstable outcomes across different datasets, indicating a limited generalization ability. This is probably because Hin2Img and Hindroid are more dependent upon large training samples and thus has lower precision on some specific datasets. MatchGNet may have limited its performance by neglecting the correlation information between Apps during the construction of the graph. In Drebin, SVM is leveraged as the feature-based machine learning technique, making it difficult to deal with malware with rapidly changing features. DroidEvolver is also based on feature engineering and updates its model in an online manner according to out-of-sample Apps, leading to a competitive classification accuracy. Nevertheless, purely relying on explicit features is intrinsically deficient compared with semantic-rich approaches.  

\mypara{Comparison against traditional feature-based ML models.} We mainly use Random Forest (RF), Logistic Regression (LR), Decision Tree (DT), Gradient Boosting Decision Tree (GBDT) and AdaBoost as comparative baselines. In this experiment, we particularly use \textit{v2017} as the train set to build the \HIN , whilst leveraging the out-of-sample Apps with various released time or various source as the test set. Following the method in \cite{li2018significant}, we extract information from permission, API, class name, interface name and \texttt{.so} file to construct the feature vector with 63,902 dimensions, which are reduced to 128 dimensions via principal component analysis (PCA).

Fig.~\ref{machine} illustrates the F1 score and accuracy score produced by different models over different test sets. Observably, \projtitle stably outperforms all traditional baselines in all cases when carrying out the App classification. Traditional ML approaches are competitive (with Acc or F1 score around 0.95) only when the testing set is aligned with the training set (\textit{v2017}) while \projtitle can
constantly deliver precise results. Interestingly, the performance of traditional approaches is constantly poor over the dataset of some specific years, e.g., \textit{v2014} and \textit{c2019}. After examining the features involved in the PCA, we infer the root cause for this phenomenon is because some features are preferably used by malicious Apps in those years but have yet been captured in the training set. For example,  \texttt{'Ljava/lang/Cloneable'} and the .so file \texttt{'libshunpayarmeabi'} manifests in \textit{v2014} as the dominating features in the PCA but they are less important in the principle components in \textit{v2017}. Similar observations can also be found for the \textit{c2019}. This is an interesting research finding while the further in-depth study is currently beyond the scope of this paper and will be left for future work. 

To sum up, the disparity of precision implies
the difficulty in applying traditional ML models -- merely relying on explicit feature extraction -- into reliable malware detection considering the explosively growing types and numbers of Apps in the market. In comparison, \projtitle is able to mine the high-order relations between Apps, with the help of \HIN, and thus has strong generalization, i.e., high effectiveness regardless the type and size of datasets.

\vspace{-0.6mm}
\subsection{Detection Efficiency} 
\label{sec:exp:efficiency}
\vspace{-0.3mm}

\mypara{Time consumption.} In this experiment, we compare the time efficiency of our incremental detection design \modelplus against those comparative approaches with an acceptable detection accuracy (demonstrated in \S \ref{sec:exp:effectivenss}), i.e., rerunning HAN,  rerunning Metagraph2Vec, Drebin, DroidEvolve and HG2Img. It is worth mentioning that we exclude the extraction time from calculating the overall execution time for the sake of simplicity because all approaches in our experiment share the same procedure of feature extraction. In fact, it approximately takes 6.9 seconds per App to extract the feature information from its original APK file.

\begin{figure}[t]
\centerline{\includegraphics[width=0.92\columnwidth]{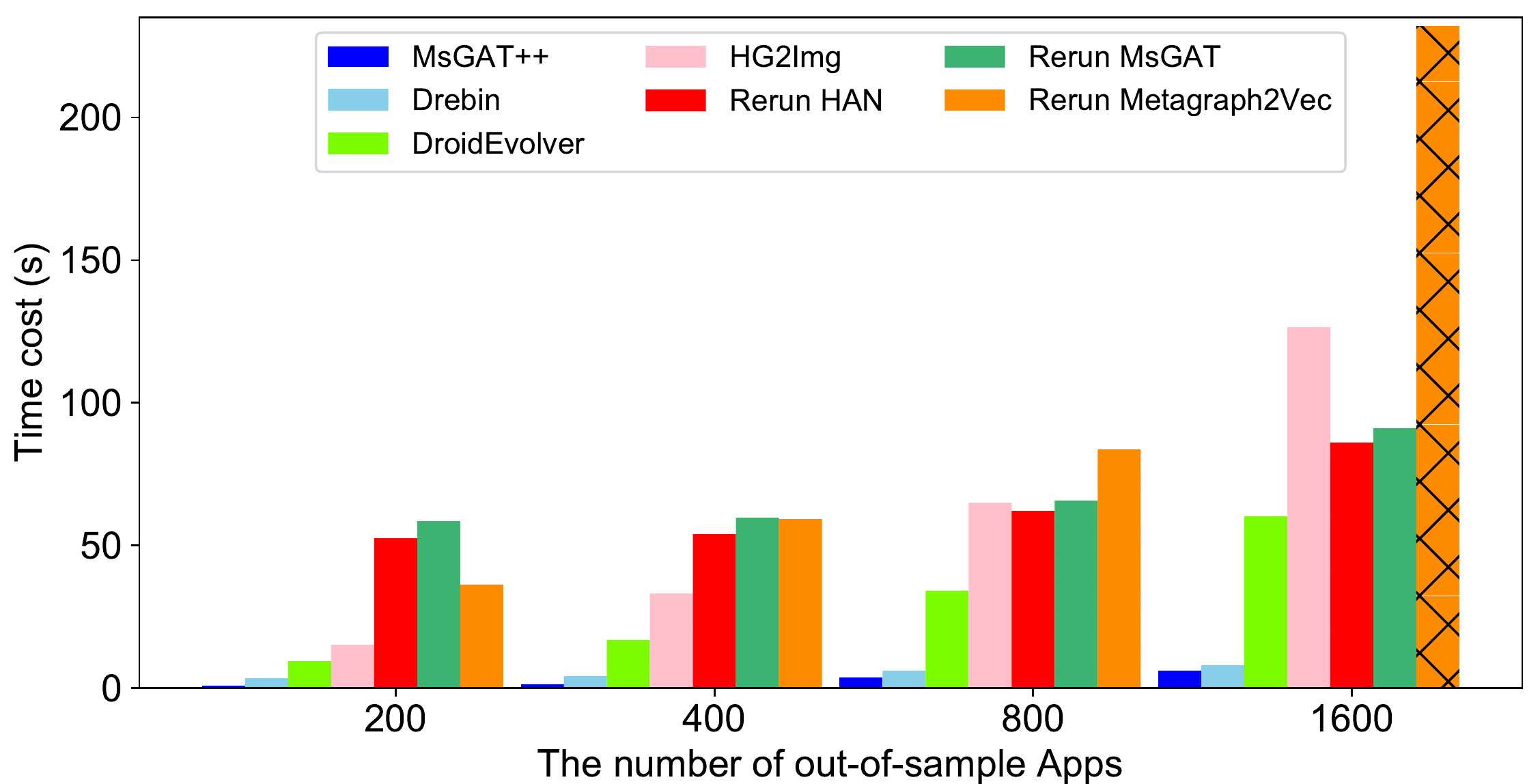}}
\vspace{-1.8mm}
\caption{Efficiency comparison of detecting out-of-sample Apps.}
\vspace{-1.2mm}
\label{fig:efficiency}
\end{figure}

As observed in Fig.~\ref{fig:efficiency}, the execution time of \modelplus is much shorter than other approaches. \modelplus takes only 3.5 milliseconds on average to detect a single out-of-sample App. This millisecond level detection by \projtitle illustrates its suitability in the real-time malware detection scenario at scale. 
In particular, \modelplus can accelerate the training time by 50× against the native approach that rebuilds the \HIN and reruns the \modelbasic. The acceleration primarily derives from our incremental learning design that can make full use of  previously learned information without the need of rerunning the entire model. In addition, \modelplus merely selects a fixed number of neighbor nodes to re-calibrate the embedding so that the time consumption only increases linearly with the increment of out-of-sample number. 

By contrast, other rerunning \HIN-based baselines is predominantly dependent upon updating embedding for all nodes based on the starting relation matrix. This leads to  discrepancies between \modelplus and others with the rerunning policy when tackling out-of-sample Apps. HG2Img relies on a certain amount of update operations to learn new features, resulting in a non-negligible time consumption. 

\mypara{System overhead.} 
Overall, the overheads are generally low, mainly generated from loading model data and carrying out the multi-tiered aggregation operations. Runtime memory consumption is typically determined by the number of nodes and features involved in the model training. The total memory consumption of \projtitle is roughly 330MB on average, far lower than the consumption of re-running based baselines (20.88GB on average). This is because all in-sample and out-of-samples have to fully loaded into memory and involved in the embedding calculation while our incremental design significantly reduce such costs.  Correspondingly, \projtitle merely uses 3.1\% additional CPU utilization on average, mainly for sorting out top-$\sigma$ samples. By contrast, the CPU utilization is up to 76\% in rerunning baselines wherein CPU-intensive matrix operations have to be performed. The low system cost also indicates the suitability of applying \projtitle into massive-scale malware detection.

\begin{table}[t]
	\centering
	\caption{Ablation Analysis}
	\label{ablation}
    \renewcommand\arraystretch{1.6}
    \footnotesize
    \scalebox{0.95}{
	\begin{tabular}{p{100pt}<{\centering}p{21pt}<{\centering}p{21pt}<{\centering}p{68pt}<{\centering}}\toprule
		Model&$Acc$&$F1$&AvgDetectionTime\\
		\hline
		\projtitle &0.9695&0.9689&3.5ms\\	
		\projtitle-I (w/o \modelbasic) &0.8731 & 0.8725 & 1.8ms\\
        \projtitle-R (w/o \modelplus) &0.9769&0.9769&205ms\\	
		\toprule
	\end{tabular}
	}
\end{table}

\vspace{-1.5mm}
\subsection{Microbenchmarking}
\label{sec:exp:microbenchmark}
\vspace{-0.6mm}

\mypara{Ablation analysis.} To investigate the impact of each component, we remove one component at a time from our model and study the individual impact on the effectiveness of detecting the out-of-sample Apps. 
We identify two tailored subsystems: i) \projtitle-I by only retaining native GAT model and removing the hierarchical GAT structure from \projtitle and ii) \projtitle-R by excluding the incremental design. Table~\ref{ablation} reports their accuracy and average time to detect a single App on \textit{v2017}. 

Without multi-step and hierarchical aggregation within a meta-structure and across meta-structures, \projtitle-I can reduce the average detection time to 1.8ms. However, both accuracy and F1 score are reduced by 9.9\% 
compared with \projtitle. This phenomenon demonstrates the accuracy gain stemming from fusing embedding results under different meta-structures. 
\projtitle-R takes far longer time to detect a malware App, simply because no incremental model is loaded and everything needs to be re-trained from  scratch. Inherently, although the accuracy experiences a negligible increase due to the full data involved in the model training, the detection efficiency of \projtitle-R is still unacceptable taking into account the long execution time. Hence, it is necessary to adopt the incremental \modelplus to ensure a reliable and rapid malware detection.
 
\begin{figure}[t]
\centerline{\includegraphics[width=0.915\columnwidth]{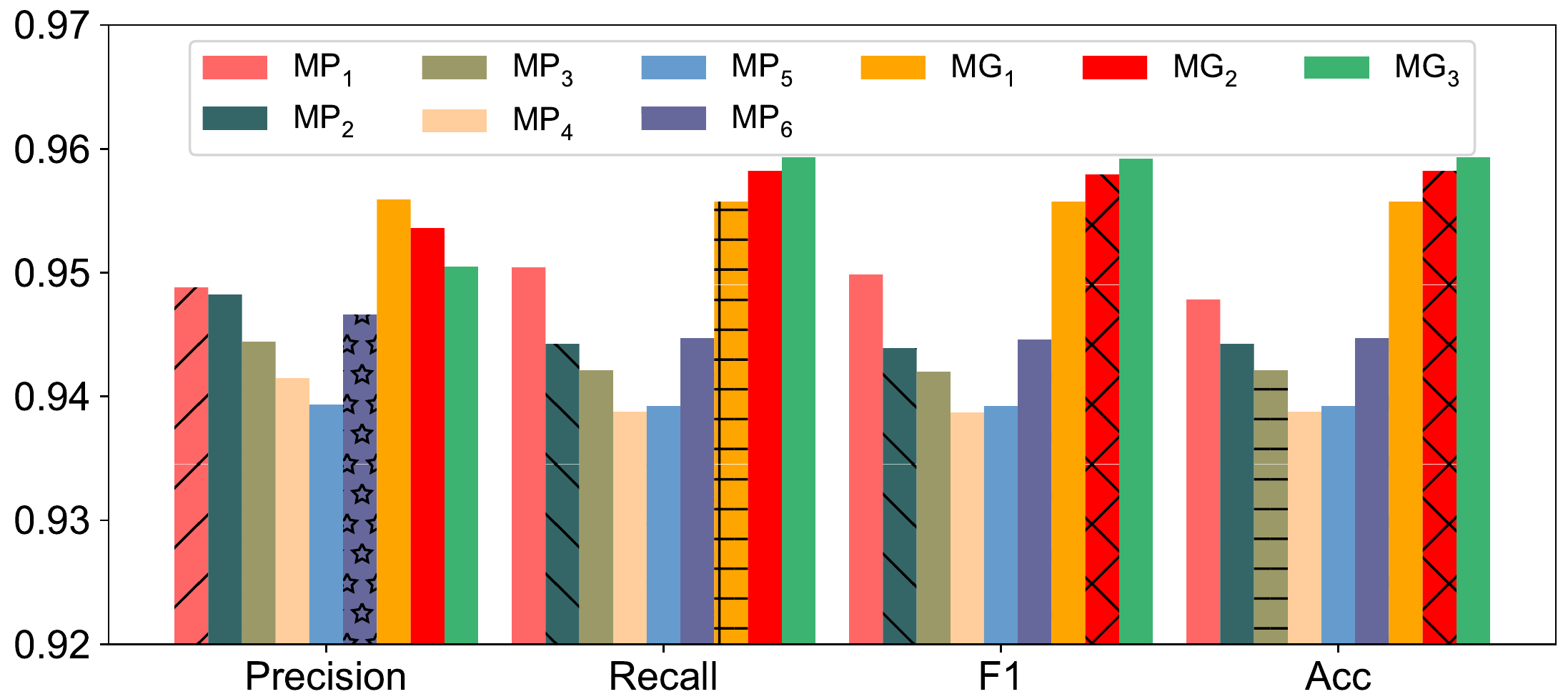}} 
\vspace{-1.3mm}
\caption{Model performance under different path combinations.}
\vspace{-1.2mm}
\label{perinsample}
\end{figure}

\begin{figure}[t]
\centerline{\includegraphics[width=0.95\columnwidth ]{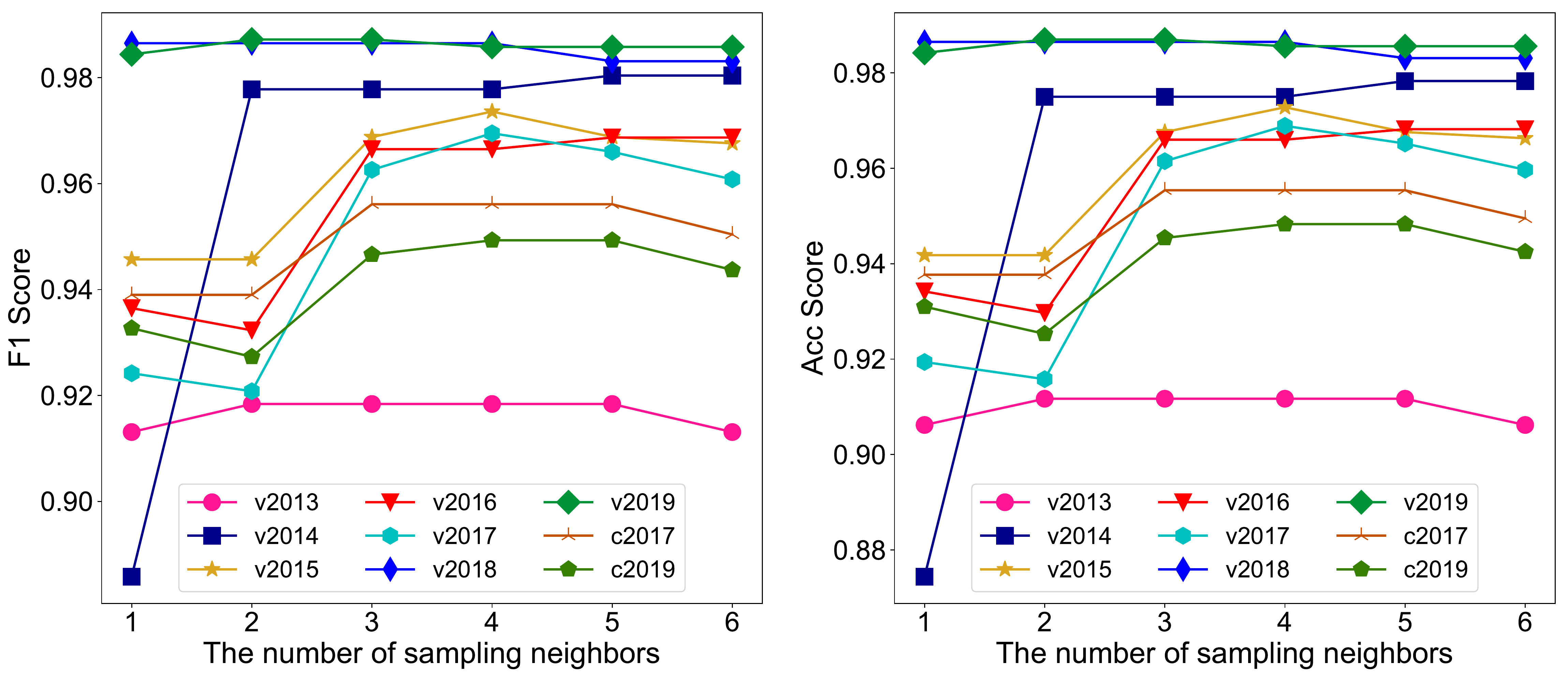}} 
\vspace{-1.3mm}
\caption{Impact of sampling neighbor number.}
\vspace{-1.2mm}
\label{neighbors}
\end{figure}

\mypara{Importance of meta-structures.}
In our model design, a group of meta-paths and meta-graphs are adopted to represent different semantic information. To ascertain the individual contribution to the detection effectiveness, we select a single meta structure at a time in this experiment.  
Fig.~\ref{perinsample} depicts the metric disparities among different meta structures. More specifically, among all meta-paths, $\mathcal{MP}_1$ and $\mathcal{MP}_4$ have the highest and lowest contribution to the detection precision. In fact, when analyzing the decompiled codes, we are able to extract far more API information than \texttt{.so} files so that the relation matrix $\mathbb{A}$ is denser than $\mathbb{S}$, and thus contains more connection information for node embedding.

Observably, using meta-graphs can achieve higher detection precision when compared to purely using meta-paths, for a combination of meta-paths can find neighbors with closer affinity. Likewise, if comparing with the results in Table~\ref{In-sample}, \modelbasic that involves the full set of semantic meta-structures unsurprisingly outperforms any situation where only a single semantic meta-structure is adopted.  This implicates that introducing sophisticated semantics is significantly meaningful to precisely uncover hidden association between entities for better classification.

\mypara{Impact of the sampling neighbor number.} As shown in Fig.~\ref{neighbors}, 
the precision  will first pick up 
within a certain range but descend once the number of sampling neighbors becomes larger (surpassing four in our experiment setting). 
In effect, increasing neighbors can provide more relevant and informative
embedding for the reference of the new nodes. However, as the neighbors begin to accumulate, noises generated by more irrelevant neighbors will, in turn, negatively impact the embedding aggregation, i.e., diminishing the representation learning effectiveness.  This implication reveals that gauging an appropriate number of neighbors is very critical to the holistic performance of embedding incoming Apps and identifying their types. We choose 3 to 4 neighbors to generate a good enough effectiveness, but one can tune the number either manually according to specific datasets or automatically empowered by reinforcement learning. This is currently beyond the scope of this paper and will be left for future work. 

\mypara{A case study of True Negative detection.}
The experiments also reveal that the true negative result manifests occasionally. In other words, a small minority of malicious Apps may not be correctly identified by our model. For example, \texttt{VirusShare_ecc4c2e7}, \texttt{VirusShare_f21ff00cf} in \textit{v2013} bypass our detection. An in-depth investigation ascertains that the embedding of such malicious apps will be assimilated by its benign neighbor nodes which are overwhelming in the process of \modelplus. In fact, since these malicious Apps have far fewer entities (no more than 30 entities) than others (normally with more than 200 entities) used in the training, the neighbors of these malicious 
apps obtained by \projtitle are sparser and tend to be benign Apps, 
resulting in the inaccurate classification. To address this problem, we plan to employ a label-aware neighbor similarity measure based on node attribute to better navigate the neighbor selection and distinguish the malware more efficiently in the future.  Nevertheless, \projtitle can achieve better detection accuracy against the up-to-date baselines, with far lower time consumption, particularly when detecting the out-of-sample Apps.

\vspace{-0.5mm}
\section{Discussion}

\mypara{Interpretablity.} \projtitle is a data-driven modeling and detecting mechanism based on Heterogeneous Information Network and network representation model empowered by Graph Attention Networks (GATs). The model's interpretability can be significantly enhanced due to the inherent nature of rich semantics, stemming from the combinations of meta-paths and meta-graphs, in the \HIN and the multi-tiered aggregation of attention from different semantics. Such an approach intrinsically outperforms the SVM based approaches such as Drebin~\cite{arp2014drebin} and Random Forest based approaches such as MaMaDroid~\cite{mariconti2016mamadroid} which has inadequate interpretability.
 
\mypara{Scalability.} The current HIN-based data modeling is scalable and can be easily extended, to any arbitrary entities and relationships, as long as the semantics can be demonstrated beneficial to the process of detection, either by domain knowledge or experimental assessment. 
In addition, since our design does not require any model rerun, the scalability can be inherently guaranteed when coping with sizable samples.  

\mypara{Robustness to obfuscation.} The semantic meta-structures based on multiple entities -  including permission, permission types, classes, interfaces, etc. - can overcome the inefficiency of API-alone detection approaches and provide a robust and accurate mechanism for detecting potential malware, in the face of API obfuscation, packing, or dataset skew (e.g., samples with less visible features such as .so files in the dataset $v2013$).  Particularly, the multi-tiered attention aggregation can automatically set the weight of different meta-paths or meta-graphs, thereby substantially reducing the impact of a single factor, e.g., the API obfuscation, on the numerical embedding and increasing the capability of generalization over different datasets and scenarios.   

\mypara{Model aging and decays.} 
Concept drift (aka. model aging, model decays) usually makes trained models fail to function on new testing samples, primarily due to the changed statistical properties of samples over time. 
The existing work \cite{zhang2020enhancing, pendlebury2019tesseract, jordaney2017transcend} measured how a model performs over time facing the concept drift, underpinned the root causes for such drift and proposed enhanced approaches to improve the model sustainability. However, active learning typically involves massive labeling for tens of thousands of malware samples, usually at a significant cost of human efforts. By far, this issue is not the focus and objective of \projtitle; In contrast, \modelplus in \projtitle aims to rapidly embed and detect the out-of-sample Apps, based upon the existing embedding results, assuming a relatively stable statistical characteristics of the existing Apps. At present, model evolving will be carried out through rerunning of \modelbasic, which is demonstrated acceptable in terms of accuracy and time consumption (detailed in \S\ref{sec:exp:efficiency}). More advanced mechanism for improving the model evolution will be left for the future work.

\vspace{-1.2mm}
\section{Related Work}
\label{sec:secRework}
\vspace{-0.3mm}

\mypara{Malware detection based on traditional feature engineering.}
Feature engineering and machine learning based  malware detection methods are two-fold: static/dynamic feature analysis. 
Static features analysis approaches \cite{li2018significant,hou2016droiddelver,2mclaughlin2017deep,arp2014drebin,xu2019droidevolver,mariconti2016mamadroid} typically include features including permissions, signatures, API sequences, etc. and directly employ such machine learning models as Random Forest, SVM or CNN for malware detection. However, they inevitably over-assume that all behaviors reflected by features should be involved within the model training, thereby having inadequate capability of tackling unknown out-of-sample cases and causing much higher false positive ~\cite{li2018significant}. Meanwhile, cunning developers can also use obfuscation techniques to hide the malicious codes \cite{alzaylaee2020dl} or perform repackaging attacks \cite{tian2017detection} to bypass detection. \cite{xu2019droidevolver} can automatically and continually update itself when detecting malware without any human involvement. Nevertheless, this scheme only proves that it has ability to adapt to updates, but does not show its compatibility with previous data sets.
In comparison, dynamic feature analysis rely on behavior detection at runtime. Specifically, \cite{dimjavsevic2015android,hou2016deep4maldroid} extract Linux kernel system calls from Apps executed in Genymotion (Android Virtual Machine) while log analysis ~\cite{alzaylaee2020dl,tobiyama2016malware} and traffic analysis ~\cite{li2016droidclassifier, wang2020deep} facilitate to capture Apps' real-world behavior. However, it is time-consuming and unrealistic to be applied in malware detection at scale. Other models from natural language processing and image recognition can be customized and re-used in malware detection. \cite{2mclaughlin2017deep} uses a deep convolutional neural network (CNN) to analyze raw opcode sequence. \cite{vinayakumar2017deep} transforms sequences of Android permissions into features by using LSTM layer and uses non-linear activation function for classification. \cite{xiao2019android} exploits LSTM to investigate potential relationships from system call sequences before classification. 
However, since Apps are constantly updated, explicit features extraction from limited Apps is ineffective in detecting unseen Apps. 

\mypara{Malware detection based on graph networks.} Gotcha~\cite{fan2018gotcha} builds up a \HIN and utilizes meta-graph based approach to depict the relevance over PE files,
which captures both content- and relation-based features of windows malware. 
HinDroid~\cite{3hou2017hindroid} is primarily on the basis of a \HIN built upon relationships between APIs and Apps, and employs multi-kernel SVM for software classification. MatchGNet~\cite{mgnet} combines \HIN model with GCN \cite{kipf2016semi} to learn graph representation and node similarity based on the invariant graph modeling of the program's execution behaviors.  \cite{wang2019attentional} constructs heterogeneous  program behavior graph, particularly for IT/OT systems, and then introduces graph attention mechanism \cite{vaswani2017attention} to aggregate information learned through GCN on different semantic paths with weights. 
However, all these methods are impeded by the static nature of the heterogeneous information network, i.e., they have limited capability of tackling emerging Apps outside the constructed graph. AiDroid~\cite{ye2018aidroid} represents each out-of-sample App with CNN~\cite{simonyan2014very}. However, the non-negligible time inefficiency stemming from multiple convolution operations becomes a potential bottleneck. \projtitle presents the first attempt to bridge the HIN-based embedding model and graph attention network to underpin incremental and rapid malware detection particularly for out-of-sample Apps.

\vspace{-0.15mm}
\section{Conclusion and future work}
\label{conclusion}
\vspace{-0.3mm}

Malware detection is a critical but non-trivial task particularly in the face of  ubiquitous Android applications and the increasingly intricate malware. In this paper, we propose \projtitle, an Android malware detection framework to rapidly and incrementally learn and identify new Android Apps. \projtitle presents the first attempt to marry the \HIN-based embedding model with graph attention network (GAT) to obtain the numerical representation of Android Apps so that any classifier can easily catch the malicious ones.
Particularly, we exploit both meta-path  and  meta-graph to best capture the implicit higher-order relationships among entities in the \HIN.  
Two learning models, \modelbasic and incremental \modelplus, are devised to fuse neighbors' embedding within any meta-structure and across different meta-structures and pinpoint the proximity between a new App and existing in-sample Apps. Through the incremental representation learning model, \projtitle can  carry out malware detection dynamically for  emerging Android Apps. Experiments show \projtitle outperforms all baselines in terms of accuracy and time efficiency. 
In the future, we plan to integrate \projtitle to smart mobile devices by devising lightweight and efficient graph convolution models, such as \cite{FGAT,LI2020188} to replace the existing modules. We also plan to investigate more advanced mechanism for underpinning the model evolving in the face of model decays particularly in federated learning environments.  

\vspace{-1.2mm}
\bibliographystyle{IEEEtran}
\bibliography{mybib}

\vskip -2.8\baselineskip
\begin{IEEEbiography}[{\includegraphics[width=1in,height=1.25in,clip,keepaspectratio]{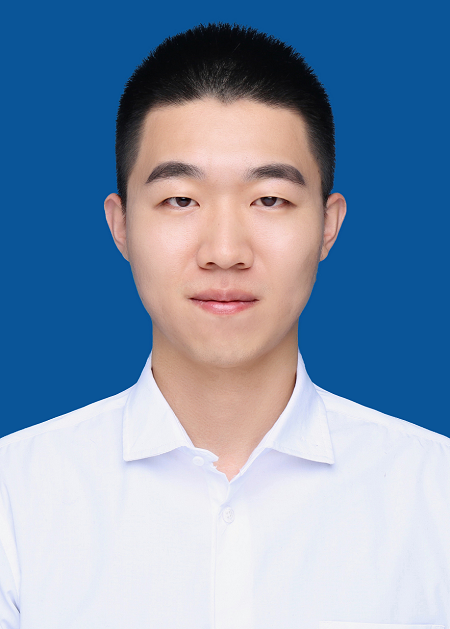}}] {Yiming He} is a PhD student at the School of Cyber Science and Technology in Beihang University, Beijing, China. His research interests include deep learning, information security and applied cryptography.
\end{IEEEbiography}
\vskip -3.6\baselineskip

\begin{IEEEbiography}[{\includegraphics[width=1in,height=1.25in,clip,keepaspectratio]{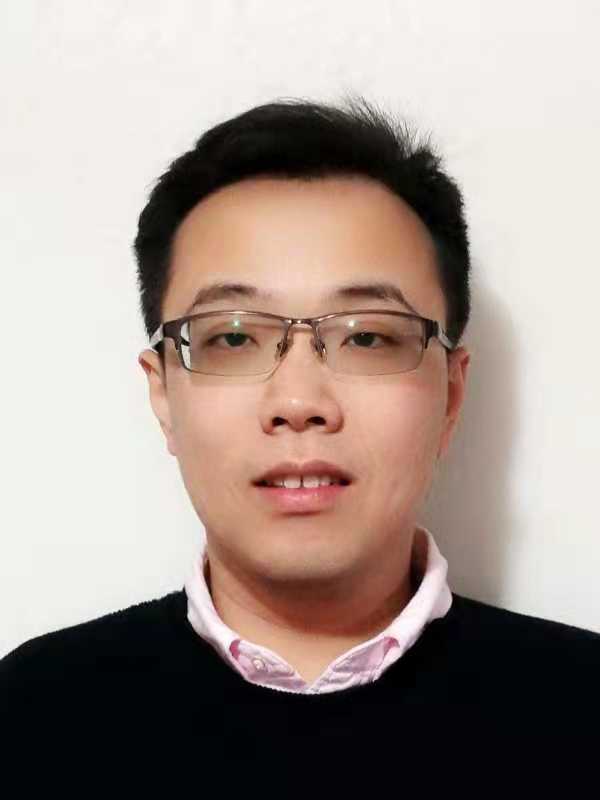}}] {Renyu Yang} is an EPSRC-funded Research Fellow with the University of Leeds, UK. He was previously with BDBC Reseach Center China, Alibaba Group China and Edgetic Ltd. UK, having industrial experience in building large-scale distributed systems with ML and co-authored/co-led many research grants including UK EPSRC, Innovate UK, EU Horizon 2020, etc. His research interests include distributed systems, resource management and applied machine learning. He is a member of IEEE.
\end{IEEEbiography}

\vskip -1.5\baselineskip
\begin{IEEEbiography}
[{\includegraphics[width=1in,height=1.25in,clip,keepaspectratio]{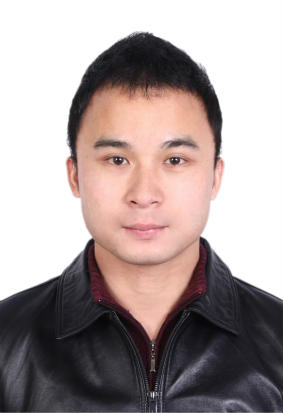}}]
{Hao Peng} is currently an Assistant Professor at the School of Cyber Science and Technology, and Beijing Advanced Innovation Center for Big Data and Brain Computing in Beihang University. His research interests include representation learning, machine learning and graph mining.
\end{IEEEbiography}
\vskip -4\baselineskip

\begin{IEEEbiography}[{\includegraphics[width=1in,height=1.25in,clip,keepaspectratio]{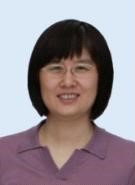}}] 
{Lihong Wang} is a professor in National Computer Network Emergency Response Technical Team/Coordination Center of China. Her current research interests include information security, cloud computing, big data mining and analytics, Information retrieval and data mining. 
\end{IEEEbiography}
\vskip -4\baselineskip

\begin{IEEEbiography}[{\includegraphics[width=1in,height=1.25in,clip,keepaspectratio]{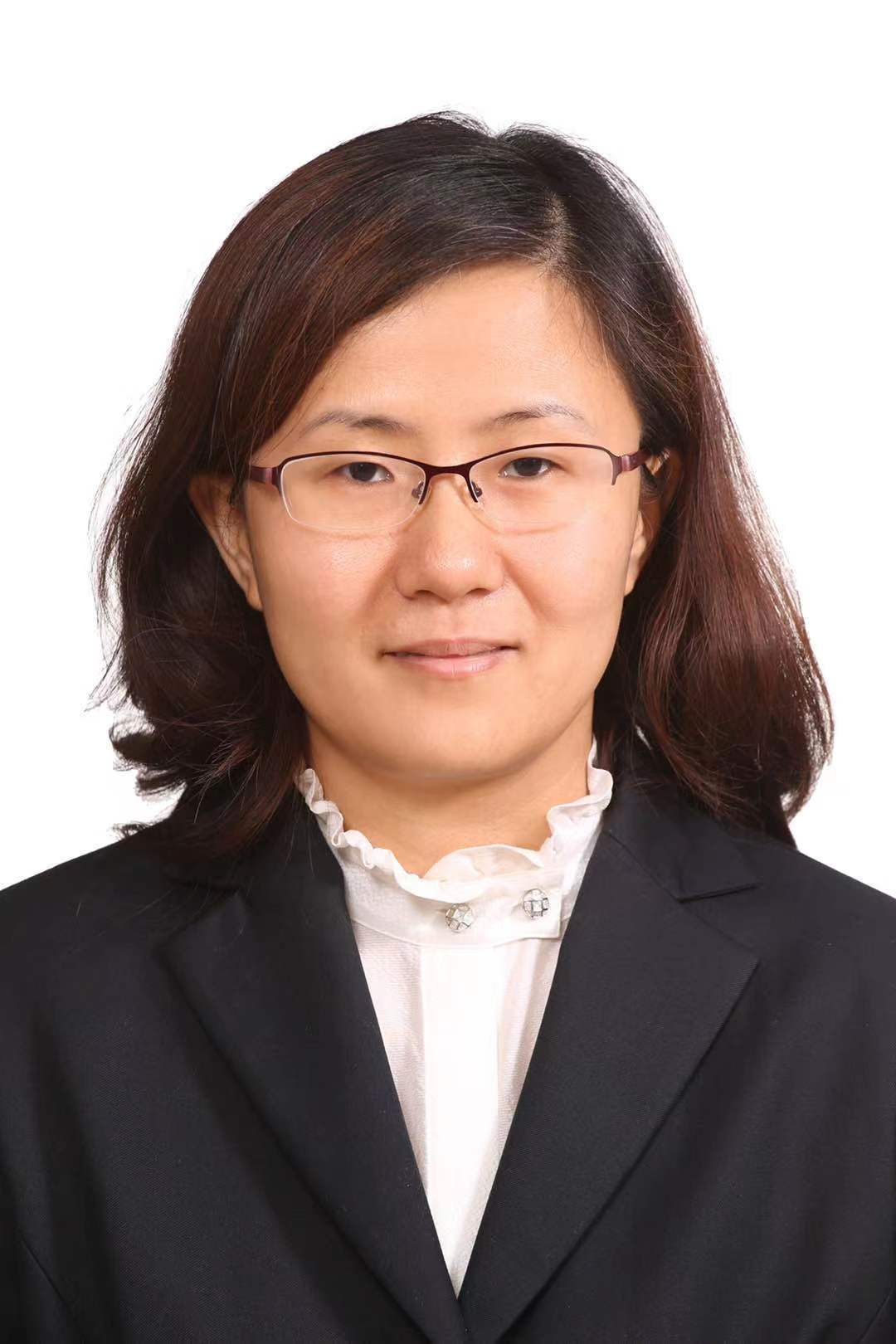}}] 
{Xiaolin Xu} is a professor in National Computer Network Emergency Response Technical Team/Coordination Center of China. Her current research interests include information security, big data mining and analytics, network security detection. 
\end{IEEEbiography}
\vskip -4\baselineskip

\begin{IEEEbiography}[{\includegraphics[width=1in,height=1.25in,clip,keepaspectratio]{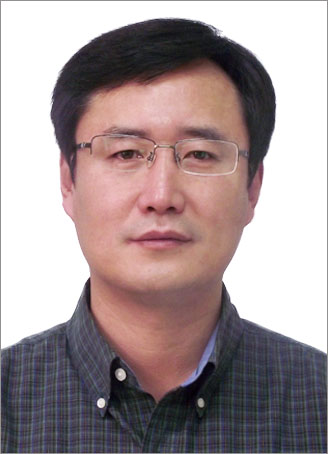}}] {Jianwei Liu} is now a professor at the School of Cyber Science and Technology in Beihang University. His current research interests include information security, communication network and cryptography.
\end{IEEEbiography}
\vskip -4\baselineskip

\begin{IEEEbiography}[{\includegraphics[width=1in,height=1.25in,clip,keepaspectratio]{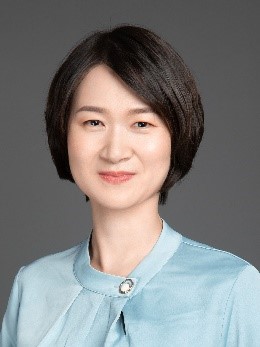}}] {Hong Liu} is currently an associate professor in East China Normal University and Shanghai Institute of Intelligent Science and Technology, Tongji University. She is also the CTO in Shanghai Trusted Industrial Control Platform Ltd. China. Her research interests include the security and privacy issues in vehicular edge computing, and industrial internet of things. She has published more than 30 SCI papers, and Google Scholar citations are 2800 times.
\end{IEEEbiography}
\vskip -4\baselineskip

\begin{IEEEbiography}[{\includegraphics[width=1.2in,height=1.25in,clip,keepaspectratio]{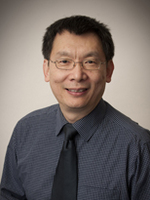}}] {Jie Xu} is the Chair Professor of Computing at University of Leeds, the leader for a Research Peak of Excellence at Leeds, Director of UK EPSRC WRG e-Science Centre, Executive Board Member of UK Computing Research Committee (UKCRC), and Chief Scientist of BDBC, Beihang University, China. He is a Steering/Executive Committee member for numerous IEEE conferences and led or co-led many research projects to the value of over \$30M, and published in excess of 400 academic papers, book chapters and edited books. His research interests include large-scale dependable distributed systems, cloud systems, big data processing, etc. He is a member of IEEE.
\end{IEEEbiography}
\vskip -4\baselineskip

\begin{IEEEbiography}[{\includegraphics[width=1in,height=1.25in,clip,keepaspectratio]{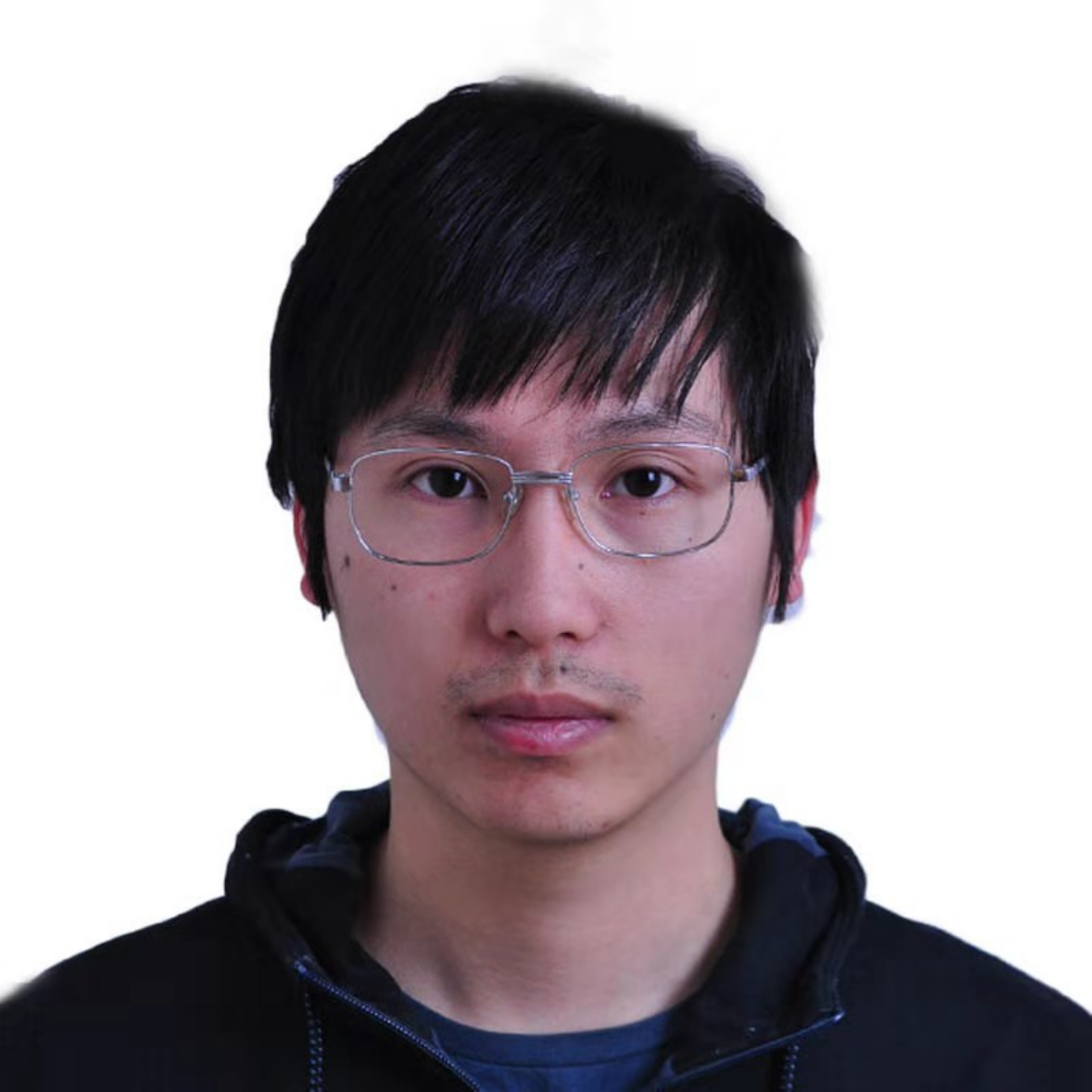}}] {Lichao Sun} is currently an Assistant Professor in Lehigh University, USA. He obtained his PhD from the University of Illinois at Chicago, US. His research interests include deep learning and data mining. He mainly focuses on security and privacy, social network and natural language processing applications. 
\end{IEEEbiography}
\vskip -3.5\baselineskip

\end{document}